\documentclass{article}
\usepackage{preprint,times}

\usepackage{amsmath}
\usepackage{fancybox}
\usepackage[most]{tcolorbox}
\usepackage{fvextra}
\usepackage{needspace}
\usepackage{booktabs}
\usepackage{tabularx}
\usepackage{array}
\usepackage{colortbl}
\usepackage{float}
\usepackage{graphicx}
\usepackage{hyperref}
\usepackage{url}

\newcolumntype{Y}{>{\centering\arraybackslash}X}
\definecolor{tblNavy}{HTML}{203348}
\definecolor{tblTrackA}{HTML}{2464B4}
\definecolor{tblTrackB}{HTML}{C74E30}
\definecolor{tblTrackC}{HTML}{087F77}
\definecolor{tblATint}{HTML}{E8F0FA}
\definecolor{tblBTint}{HTML}{FBECE6}
\definecolor{tblCTint}{HTML}{E7F4F0}
\definecolor{tblZebra}{HTML}{EFF4F7}
\definecolor{tblGroup}{HTML}{DCE6EE}
\newcommand{\tblHead}[1]{\textcolor{tblNavy}{\textbf{#1}}}
\newcommand{\tblTrackHead}[2]{\cellcolor{tbl#1Tint}\textcolor{tblTrack#1}{\textbf{#2}}}
\newcommand{\tblTrackLabel}[1]{\textcolor{tblTrack#1}{#1}}

\input{title_brand}
\newcommand{\modelicon}[2]{\raisebox{-0.2\height}{\includegraphics[height=#1]{tables/logos/#2}}}

\newcommand{\ie}{{\emph{i.e.}}}

\title{\omniwordmark{}: Benchmarking Audiovisual\protect\\ Website Development}
\hypersetup{pdftitle={Omni2Web: Benchmarking Audiovisual Website Development}}
\author{%
\begin{minipage}{\dimexpr\textwidth-2\tabcolsep\relax}
\centering
{\large
\textbf{Minghao Han}$^{1,2,*}$
\quad \textbf{Zhenghao Xing}$^{3,*}$
\quad \textbf{Xize Cheng}$^{2}$
\quad \textbf{Yuxuan Wang}$^{2}$\\[0.2em]
\textbf{Junming Lin}$^{2,4}$
\quad \textbf{Ling Wang}$^{2,5}$
\quad \textbf{Yinsong Yan}$^{2,5}$\\[0.2em]
\textbf{Yunfei Chu}$^{2}$
\quad \textbf{Qize Yang}$^{2}$
\quad \textbf{Jin Xu}$^{2,\dagger}$\\[0.9em]
}
{\normalfont
$^{1}$FDU
\quad $^{2}$Alibaba Token Hub, Alibaba Group
\quad $^{3}$CUHK
\quad $^{4}$THU
\quad $^{5}$PolyU
}
\\[0.35em]
{\normalfont\small
$^{*}$Equal contribution.
\quad $^{\dagger}$Corresponding author.
}
\\[0.5em]
{\normalfont\small
\href{https://omni2web-bench.github.io/}{%
\tikz[baseline=-0.1ex,x=1em,y=1em,line width=0.45pt]{%
\draw[rounded corners=0.8pt] (0,0) rectangle (1,0.72);
\draw (0,0.5) -- (1,0.5);
\fill (0.14,0.61) circle[radius=0.035];
\fill (0.28,0.61) circle[radius=0.035];
\fill (0.42,0.61) circle[radius=0.035];
}}\hspace{0.35em}\url{https://omni2web-bench.github.io/}}
\end{minipage}
}

\begin{document}

\maketitle
\lhead{Omni2Web: Benchmarking Audiovisual Website Development}

\begingroup
\setlength{\textfloatsep}{8pt}
\setlength{\floatsep}{10pt plus 2pt minus 2pt}

\begin{abstract}
Screen-recorded web editing requests contain weak deictic expressions such as ``this'' and ``there,'' whose referents depend on speech, cursor trajectories, page state, and edit history. Such requests require intent recovery beyond the explicit specifications assumed by many existing web-editing benchmarks. We introduce Omni2Web, a bilingual benchmark of 918 instances spanning 13,907 edit steps. It defines three complementary tracks: Direct Editing evaluates webpage editing from recordings, Instruction Recovery measures explicit intent recovery, and Instruction Utility tests whether recovered instructions can drive a fixed code executor. We evaluate 17 open- and closed-source models. The best models attain 51.17 on the Edit Fidelity Score (EFS) for Direct Editing and 49.14 on the Instruction Recovery Score (IRS); under the fixed executor, the strongest recovered instructions reach 51.08 EFS, still far below the 89.69 EFS obtained with oracle instructions. Step-level analyses show that correct grounding does not guarantee successful edits, while some Omni models recover instructions that the fixed coding model executes substantially better than their direct edits. Controlled ablations further demonstrate the value of temporally aligned audiovisual evidence, while alternative judges preserve the leader and broad ordering. Together, these findings reveal substantial headroom in multimodal intent recovery and code execution and highlight the promise of pairing Omni rewriters with coding models.
\end{abstract}

\section{Introduction}
\label{sec:introduction}

\textit{``Move this below that, make these two match, and undo the previous color change.''} For a viewer, this is actionable; as text alone, it is incomplete because cursor position, speech timing, and prior edits convey its referents. The operation spans language, cursor movement, and visual context.

Existing evaluations largely avoid this ambiguity. Instructed-editing benchmarks begin with explicit textual specifications~\citep{chi2026editbench}; web and computer-use benchmarks execute actions from explicit goals~\citep{deng2023mind2web,koh2024visualwebarena,xie2024osworld,lu2024weblinx}. GUI-centered work emphasizes screen understanding, localization, or action prediction~\citep{baechler2024screenai,hong2024cogagent}, while audio-visual benchmarks test temporal alignment without realizing the evidence as a code edit~\citep{zhou2025dailyomni}. None directly tests whether a model can recover weakly expressed intent from a recording and turn it into a complete, safe webpage modification.

A single end-to-end editing score would also obscure two distinct failures. A model may fail because it cannot determine what the user intends to change, or because it understands the request but cannot implement it correctly in HTML. The former concerns multimodal reference resolution; the latter concerns code execution. Successful editing must also preserve page content unrelated to the requested changes. Evaluation should separate intent recovery, implementation, and preservation.

We introduce \textbf{Omni2Web}, a bilingual benchmark for web editing from screen recordings with weak references. Its 918 instances span 129 source webpages, 13,907 edit steps, and 12 operation types. Each aligns speech with cursor trajectories and includes long sequences, localized modifications, and revisions. Annotations separately capture target grounding, edit fulfillment, revision handling, and damage to unrelated content. Figure~\ref{fig:benchmark-overview} illustrates the task and evaluation design.

Omni2Web defines three complementary tracks. \textbf{Track A: Direct Editing} asks a model to produce the edited webpage directly from the source HTML and recording. \textbf{Track B: Instruction Recovery} asks it to rewrite the recording into explicit targets and editing actions. \textbf{Track C: Instruction Utility} passes these recovered instructions and the source HTML to a fixed coding model. Track C therefore pairs multimodal intent recovery with code generation, testing whether a fixed coding model can implement the instructions recovered by a multimodal model. Together, the three tracks provide complementary evidence about intent recovery and execution.

We evaluate 17 open- and closed-source models. Results leave substantial headroom: the best Direct Editing Edit Fidelity Score (EFS) is 51.17, the best Instruction Recovery Score (IRS) is 49.14, and the strongest recovered instructions score 51.08 EFS with 56.95 normalized instruction utility, versus 89.69 EFS for oracle instructions (\ie, DeepSeek-v4-pro~\citep{deepseek2026deepseek}). Step-level analyses reveal two complementary patterns: correct target grounding does not always translate into successful editing, while high No-damage can reflect inaction. Track C further shows that some Omni models recover useful instructions despite weak direct editing, allowing the fixed coding model to produce substantially stronger edits. Controlled ablations support the importance of synchronized audiovisual evidence, with native audiovisual input performing best for both tested Omni models. Shifting audio out of sync with video reduces IRS by up to 13.73 points, while adding timestamps to frame-and-transcript inputs improves it by up to 14.98 points. We contribute:

\noindent$\bullet $ We introduce Omni2Web, a bilingual benchmark comprising 918 screen-recorded editing requests, 13,907 edit steps, 129 source webpages, and 12 operation types, with synchronized audiovisual evidence and revision-aware final-state rubrics.

\noindent$\bullet $ We develop a three-track protocol that evaluates direct editing, component-level instruction recovery, and the downstream utility of recovered instructions under a fixed executor, while separately measuring preservation of unrelated page content.

\noindent$\bullet $ We benchmark 17 open- and closed-source models and conduct controlled modality, temporal-alignment, and executor studies. The results expose distinct grounding and execution bottlenecks, demonstrate the value of synchronized audiovisual evidence, and test evaluator reliability through alternative judges and human calibration.

\begin{figure}[t]
  \centering
  \includegraphics[width=\textwidth]{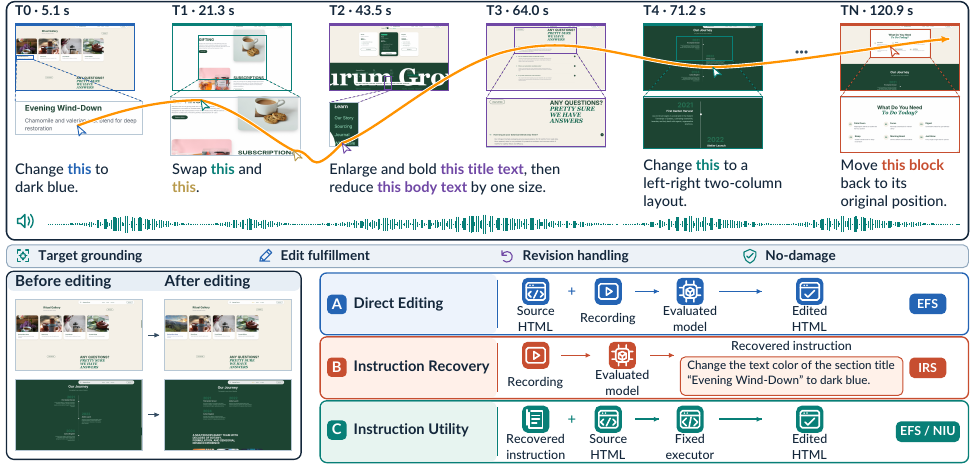}
  \vspace{-0.7em}
  \caption{Overview of Omni2Web. The orange curve shows cursor movement across the recording. Tracks~A--C evaluate direct editing, instruction recovery, and instruction utility.}
  \label{fig:benchmark-overview}
\end{figure}

\section{Related Work}
\label{sec:related-work}

\subsection{Web Editing and GUI Benchmarks}

Web agents and instructed-editing benchmarks generally assume explicit textual goals~\citep{deng2023mind2web,koh2024visualwebarena,chi2026editbench,dang2025envisioning}. Video-based benchmarks differ in output: WebVR and IWR-Bench reconstruct complete webpages~\citep{dai2026webvr,chen2026iwr}, VideoGUI predicts GUI actions~\citep{lin2024videogui}, and Frontalk studies iterative front-end development with multimodal feedback~\citep{wu2025frontalk}. Omni2Web reuses 129 WebVR source pages but collects new editing recordings and annotations. Starting from existing HTML, models must resolve weak speech--cursor references, apply multi-step revisions, and preserve unrelated content.

\subsection{Multimodal Understanding and Grounded Interaction}

Joint speech and pointing dates to ``Put-That-There,'' which used gesture to resolve deictic expressions; later work established complementary roles for language, gesture, and visual context~\citep{bolt1980put,oviatt1999ten}. Modern work studies audio-visual reasoning~\citep{goel2026mmou,li2026omnivideobench,tao2026lvomnibench,chao2026jointavbench}, embodied and video grounding~\citep{shi2022spatial,ding2023mevis,wang2024grounded}, and GUI localization~\citep{baechler2024screenai,hong2024cogagent,cheng2024seeclick,you2024ferret}. These tasks output answers, locations, masks, or actions; Omni2Web requires recovered targets, actions, and revisions to yield final HTML, making multimodal grounding upstream to code editing.

\section{Omni2Web}
\label{sec:benchmark}

\subsection{Task Instances and Annotations}
\label{sec:benchmark-instances}

Omni2Web evaluates single-turn, multi-step edits to an existing webpage. Each instance comprises the source page, synchronized media, transcript, and rubric:
{\small
\begin{gather}
  \mathcal{X}_i=\left(H_i,V_i^{\mathrm{av}},V_i^{\mathrm{sil}},U_i,\tau_i,\mathcal{B}_i\right),
\end{gather}}%
where $H_i$ is the source HTML, $V_i^{\mathrm{av}}$ the original audiovisual screen recording, $V_i^{\mathrm{sil}}$ its silent-video counterpart, $U_i$ the isolated audio, $\tau_i$ the manually written transcript, and $\mathcal{B}_i$ the step-level rubric. Auxiliary metadata is omitted from the notation. The model must use the spoken action together with the visual context to resolve expressions such as ``this section'' or ``these two blocks,'' return a complete edited HTML document, and preserve unrelated parts of the source page.

Annotations decompose every instance into an ordered edit sequence. Each step has an operation type and two rubric items: one checks target grounding and the other checks edit fulfillment (or revision handling for a revision step). An instance-level no-damage item checks whether content outside the requested edits remains intact. The explicit reference instructions $I_i^*$ are constructed deterministically from the paired rubric anchors in $\mathcal{B}_i$. We use $M_i$ for the model-facing input assembled from $V_i^{\mathrm{av}}$, $V_i^{\mathrm{sil}}$, and $\tau_i$ under the protocol in Section~\ref{sec:input-construction}.

\subsection{Dataset Construction}
\label{sec:dataset-construction}

We select 129 source webpages from WebVR~\citep{dai2026webvr}. Annotators write multiple edit scripts per page, collectively covering the 12 operation types in Figure~\ref{fig:benchmark-anatomy}(a). The scripts specify the requested changes and their order. We review them before recording. Annotators then describe the requests naturally while pointing to the relevant page elements with the cursor. Weak references such as ``here'' and ``these two'' depend on speech, cursor pointing, and page context for their meaning. Annotators also manually write the transcripts $\tau_i$ and rubric anchors. The transcripts capture the spoken requests. The anchors specify each step's target and editing requirements. Before inclusion, we review the edit sequences and reference annotations for consistency in targets and edits.

A professional external team of 15 paid annotators collected and annotated the data. All received standardized training and provided informed consent for research use and public data release.

\subsection{Text-Only Recoverability Probe}
\label{sec:difficulty-probe}

Some deictic requests remain recoverable without video when $(\tau_i,H_i)$ identifies a plausible target. We therefore measure text-only recoverability with a separate diagnostic probe. Gemini 3.1 Pro~\citep{google2026gemini31pro} receives $(\tau_i,H_i)$ but no video or rubric anchors and predicts one target description per step. GPT-5.5~\citep{openai2026gpt55} judges whether each prediction denotes the same page element or region as the corresponding human target anchor. Missing predictions score zero. This fixed evaluation pipeline is applied to all instances.

For instance $i$ with $S_i$ steps, the probe hit rate is
{\small
\begin{gather}
  h_i=\frac{1}{S_i}\sum_{j=1}^{S_i} \mathbf{1}\!\left[\widehat{t}_{ij}\equiv t_{ij}\right],
\end{gather}}%
where $\widehat{t}_{ij}$ is the predicted target and $t_{ij}$ is the human target anchor. We define Hard as $h_i<0.2$, Medium as $0.2\leq h_i<0.5$, and Easy as $h_i\geq0.5$. Low $h_i$ means few recovered targets. These bands are diagnostic strata, not benchmark scores or measures of intrinsic human difficulty.

\subsection{Dataset Statistics}
\label{sec:dataset-statistics}

Figure~\ref{fig:benchmark-anatomy} summarizes 918 instances from 129 source webpages, covering 13,907 historical edit steps across 12 operation types. Sequences contain 10--20 steps, with a mean of 15.15. Every instance includes one to three revisions, totaling 2,169 revision-category steps. Their 2,525 undo links span a median of two steps (Figure~\ref{fig:benchmark-anatomy}(b)). Spoken languages are nearly balanced (465 Chinese, 453 English). The median video duration is 128.04 seconds. Under the text-only probe in Section~\ref{sec:difficulty-probe}, 481 instances are Hard, 230 are Medium, and 207 are Easy. Thus, 77.45\% fall below a 0.5 recovery rate without video, indicating substantial dependence on visual and deictic evidence.

\begin{figure}[t]
  \centering
  \includegraphics[width=\textwidth]{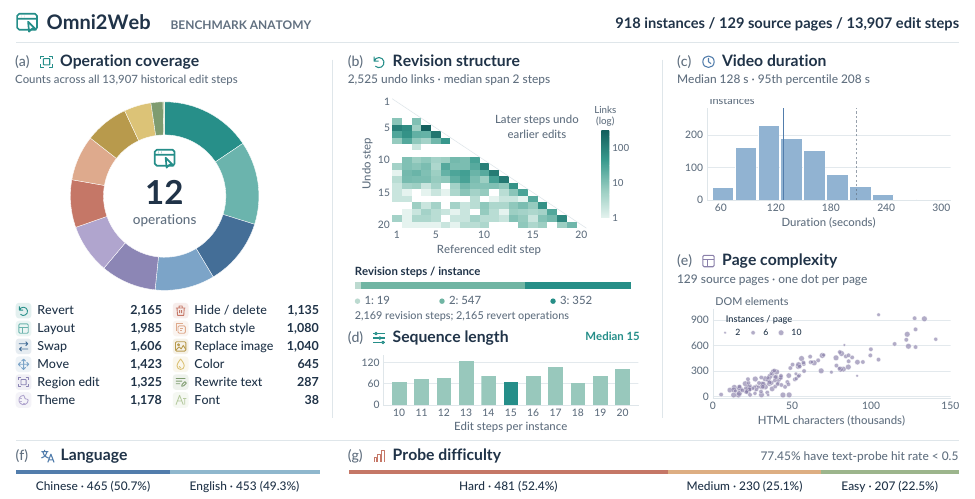}
  \vspace{-0.7em}
  \caption{Dataset anatomy of Omni2Web. In (b), colors show link counts on a log scale; a revision can undo multiple steps. In (e), bubble area represents instances per source page.}
  \label{fig:benchmark-anatomy}
\end{figure}

\section{Evaluation Protocol}
\label{sec:evaluation-protocol}

\subsection{Three Evaluation Tracks}
\label{sec:evaluation-tracks}

With the notation of Section~\ref{sec:benchmark-instances}, let $f_\theta^A$ and $f_\theta^B$ denote the evaluated model prompted for Tracks A and B, respectively, and $e$ the fixed DeepSeek-v4-pro~\citep{deepseek2026deepseek} executor. The tracks are
{\small
\begin{gather}
\begin{aligned}
  \text{Track A:}\quad &\widehat{H}_i^A=f_\theta^A(H_i,M_i), && \text{direct editing}, \\
  \text{Track B:}\quad &\widehat{I}_i=f_\theta^B(M_i), && \text{instruction recovery}, \\
  \text{Track C:}\quad &\widehat{H}_i^C=e(H_i,\widehat{I}_i),\quad H_i^{\mathrm{oracle}}=e(H_i,I_i^*), && \text{instruction utility}.
\end{aligned}
\end{gather}}%
Track A produces an edited page, Track B recovers instructions without source HTML, and Track C measures their utility with a fixed executor that receives no media; the three contexts remain distinct.

\subsection{Track-Specific Scoring}
\label{sec:scoring-pipeline}
\label{sec:scoring-ac}

\vspace{0.5pt}\noindent\textbf{Tracks A\&C.} Tracks A and C share a media-free, rubric-anchored LLM judge~\citep{zheng2023judging}. Before aggregation, 2,525 earlier steps explicitly cancelled by later whole-step revisions are removed from the 13,907-step history, leaving 11,382 active steps; the cancelling revisions remain scored, while Track~B retains the full history.

Given source $H_i$, prediction $\widehat{H}_i$, and rubric $\mathcal{B}_i$, the judge returns binary decisions for active target-grounding, edit-fulfillment, and revision items. Omissions score zero. An image check can only overturn passed replace-image items. Category averages give scores $G_i$, $E_i$, and $R_i$.

Separately, a structured HTML diff compares $H_i$ with $\widehat{H}_i$. Let $\Delta_i$ be the extracted changes and $\mathcal{D}_i$ the collateral changes identified by the damage judge. Each damage receives a low, medium, or high penalty $w(d)\in\{0.05,0.15,0.30\}$. The No-damage score and EFS are
{\small
\begin{gather}
  N_i = \max\!\left(0,1-\frac{\sum_{d\in\mathcal{D}_i}w(d)}{\max(\alpha|\Delta_i|,1)}\right),
  \\
  \mathrm{EFS}_i = 0.10G_i+0.50E_i+0.20N_i+0.20R_i.
\end{gather}}%
Here $|\Delta_i|$ is the number of extracted changes. We set the No-damage normalization coefficient to $\alpha=0.15$, matching the medium penalty, while the unit floor stabilizes short or empty diffs. Category scores are averaged within each instance and EFS across instances. For Track C, instruction utility is normalized against the oracle executor run:
{\small
\begin{gather}
  \mathrm{NIU}=\frac{\overline{\mathrm{EFS}}_{\mathrm{recovered}}}{\overline{\mathrm{EFS}}_{\mathrm{oracle}}}.
\end{gather}}%
NIU is a ratio of dataset-level aggregates and is not clipped.

\vspace{0.5pt}\noindent\textbf{Track B.}\label{sec:scoring-b} The model outputs an ordered JSON list of target/instruction pairs. One-to-one matching with $I_i^*$ assigns binary target and action credit $(\pi_{ij}^{T},\pi_{ij}^{A})\in\{0,1\}^2$. Let $n_i^{\mathrm{ref}}=|I_i^*|$, $n_i^{\mathrm{pred}}=|\widehat{I}_i|$, and $m_i^c=\sum_j\pi_{ij}^c$ for $c\in\{T,A\}$. Component F1 and the Instruction Recovery Score (IRS) are
{\small
\begin{gather}
  F_i^c=\frac{2m_i^c}{n_i^{\mathrm{ref}}+n_i^{\mathrm{pred}}},
  \quad c\in\{T,A\}, \\
  \mathrm{IRS}_i = 0.50F_i^T+0.50F_i^A.
\end{gather}}%
The semantic judge uses $H_i$ only to verify element equivalence. Deterministic checks enforce one-to-one matches and valid indices. Duplicate, unmatched, and extra predictions increase $n_i^{\mathrm{pred}}$, penalizing unsupported enumeration. Missing references and unusable outputs score zero. IRS is macro-averaged over instances. Across sensitivity tests, the Track~A and C leaders persist under all 156 EFS configurations, while alternative IRS formulations retain $\rho\geq0.983$ but can change the Track~B leader (Appendices~\ref{app:efs-sensitivity} and~\ref{app:track-b-sensitivity}). Gemini 3.1 Pro judges all model-based Track~A--C stages with fixed English prompts across languages and model families. Appendix~\ref{app:prompts} lists them.

\subsection{Models and Input Construction}
\label{sec:evaluated-models}

We evaluate 17 open- and closed-source models spanning omni-modal and vision-language (VL) input interfaces. The open-source group comprises MiniCPM-o 4.5~\citep{cui2026minicpmo45}, Gemma 4 12B Unified~\citep{team2026gemma4}, Qwen3.8-27B~\citep{qwen2026qwen38}, Nemotron 3 Nano Omni~\citep{deshmukh2026nemotron}, MiMo-V2.5~\citep{xiaomi2026mimov25}, MiniMax-M3~\citep{minimax2026m3}, Qwen3.8-Max~\citep{qwen2026qwen38}, and Kimi-K3~\citep{team2026kimik3}.
The closed-source group comprises Muse Spark 1.1~\citep{menghini2026muse}, Seed2.0 Lite~\citep{bytedance2026seed20}, Gemini 3.5 Flash~\citep{google2026gemini35flash}, Qwen3.5-Omni-Plus~\citep{team2026qwen35omni}, Seed2.1 Pro~\citep{bytedance2026seed21}, Grok 4.6~\citep{xaigrok46}, Gemini 3.1 Pro~\citep{google2026gemini31pro}, GPT-5.6 Sol~\citep{openai2026gpt56}, and Claude Opus 5~\citep{anthropic2026claudeopus5}. Table~\ref{tab:main-ab-results} marks each model's input modality. Cross-family results therefore compare end-to-end systems under native input interfaces, rather than architectures under identical information.

\label{sec:input-construction}

We assign each model one main input condition based on observed media support and use it consistently on Tracks A and B. Let $F_i=\operatorname{FPS}_{1}(V_i^{\mathrm{sil}})$ denote the timestamped frame sequence sampled once per second. The model-facing input is
{\small
\begin{gather}
  M_i=\begin{cases}
    V_i^{\mathrm{av}}, & \text{Omni}, \\
    (V_i^{\mathrm{sil}},\tau_i), & \text{VL with native-video delivery}, \\
    (F_i,\tau_i), & \text{VL with frame delivery}.
  \end{cases}
\end{gather}}%

\section{Main Results}
\label{sec:main-results}

\subsection{Main Leaderboard}
\label{sec:main-leaderboard}

Tables~\ref{tab:main-ab-results} and~\ref{tab:main-c-results} report performance on all three tracks for the 918 instances. Track~C reports the absolute Edit Fidelity Score (EFS) and normalized instruction utility (NIU), normalized by the 89.69 oracle-executor EFS. Higher values are better. MiniCPM-o 4.5 and Gemma 4 12B lack Track~A scores because the direct-editing input exceeds their context limits.

Gemini 3.5 Flash leads Direct Editing with 51.17 EFS, whereas Qwen3.5-Omni-Plus has the highest Instruction Recovery point estimate, at 49.14 IRS, and leads Instruction Utility with 56.95 NIU. Omni models have the highest point estimates on all three tracks. Only Gemini exceeds 50 on Track~A; no model exceeds 50 IRS, and the best recovered instructions reach 51.08 EFS under the fixed executor versus 89.69 for oracle instructions, leaving substantial headroom.

\begin{table}[t]
  \centering
  \scriptsize
  \setlength{\tabcolsep}{2.2pt}
  \renewcommand{\arraystretch}{1.08}
  \caption{Main results on the 918 instances for Direct Editing (Track A) and Instruction Recovery (Track B). Track B columns are component F1 scores and their average IRS; all values are percentages, with column bests in bold.}
  \label{tab:main-ab-results}
  \arrayrulecolor{tblNavy}
  \begin{tabularx}{\textwidth}{@{}>{\raggedright\arraybackslash}p{2.45cm}c*{5}{Y}!{\hspace{2pt}{\color{tblNavy}\vrule width 0.3pt}\hspace{2pt}}*{3}{Y}@{}}
    \toprule
    & & \multicolumn{5}{>{\columncolor{tblATint}}c}{\textcolor{tblTrackA}{\textbf{Track A: Direct Editing}}} & \multicolumn{3}{>{\columncolor{tblBTint}}c}{\textcolor{tblTrackB}{\textbf{Track B: Instruction Recovery}}} \\
    \cmidrule(lr){3-7}\cmidrule(lr){8-10}
    \tblHead{Model} & \tblHead{Modality} & \tblHead{Grd.} & \tblHead{Edit} & \tblHead{No dmg.} & \tblHead{Rev.} & \tblHead{EFS} & \tblHead{Target F1} & \tblHead{Action F1} & \tblHead{IRS} \\
    \midrule
    \rowcolor{tblGroup}\multicolumn{10}{c}{\tblHead{Open-Source Models}} \\
    \modelicon{1.0em}{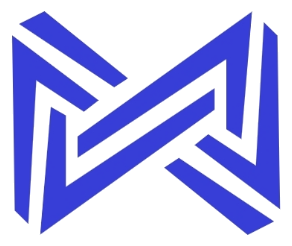}~MiniCPM-o 4.5 & Omni & -- & -- & -- & -- & -- & 17.87 & 31.93 & 24.90 \\
    \rowcolor{tblZebra}\modelicon{1.0em}{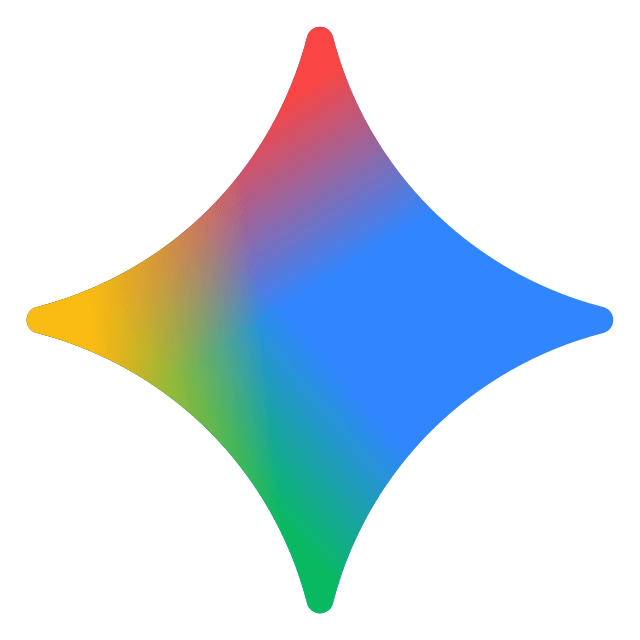}~Gemma 4 12B & Omni & -- & -- & -- & -- & -- & 1.88 & 11.67 & 6.78 \\
    \modelicon{1.0em}{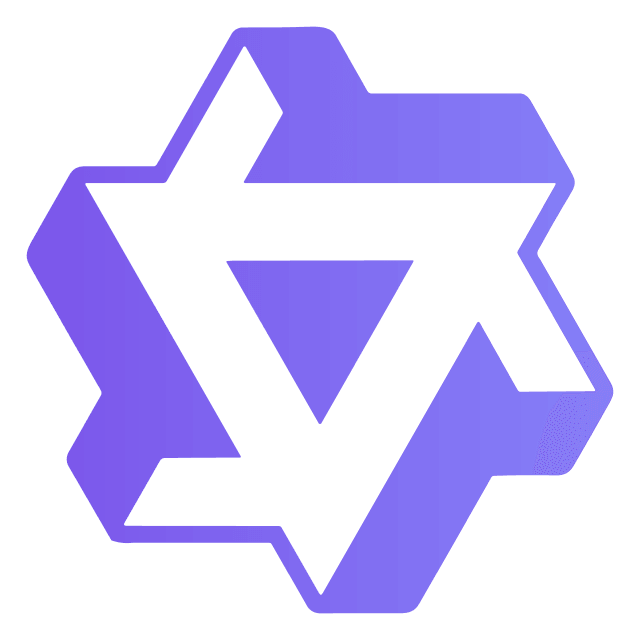}~Qwen3.8-27B & VL & 12.25 & 7.56 & \textbf{82.94} & 26.74 & 26.94 & 19.82 & 39.85 & 29.84 \\
    \rowcolor{tblZebra}\modelicon{1.0em}{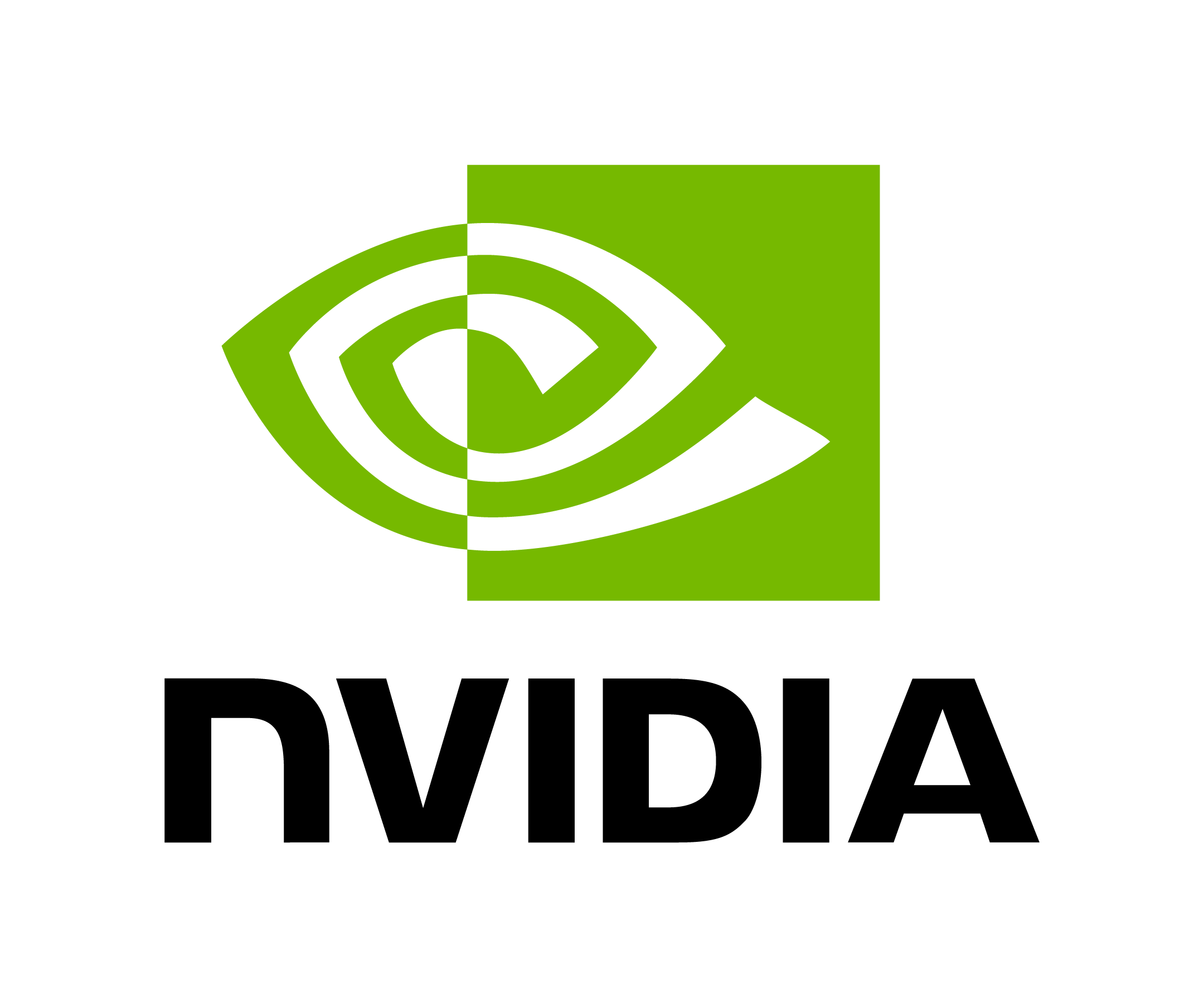}~Nemotron 3 Nano & Omni & 0.75 & 0.04 & 71.27 & 11.46 & 16.64 & 1.74 & 8.26 & 5.00 \\
    \modelicon{1.0em}{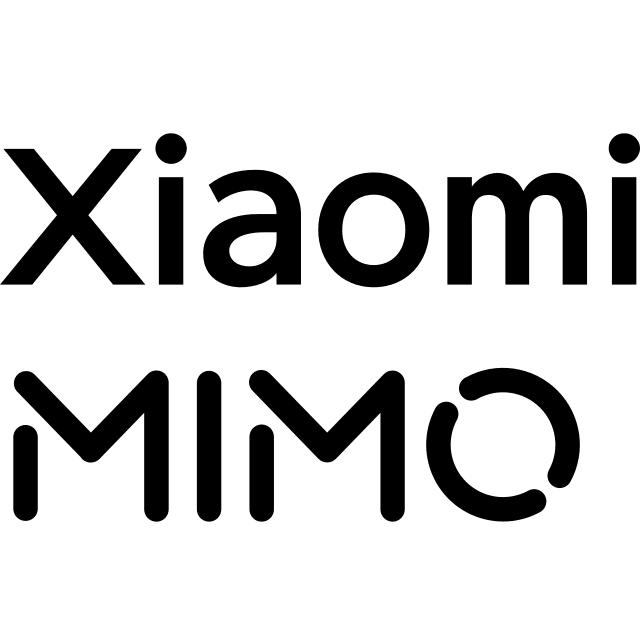}~MiMo-V2.5 & Omni & 31.44 & 17.93 & 43.59 & 39.81 & 28.79 & 33.95 & 40.67 & 37.31 \\
    \rowcolor{tblZebra}\modelicon{1.0em}{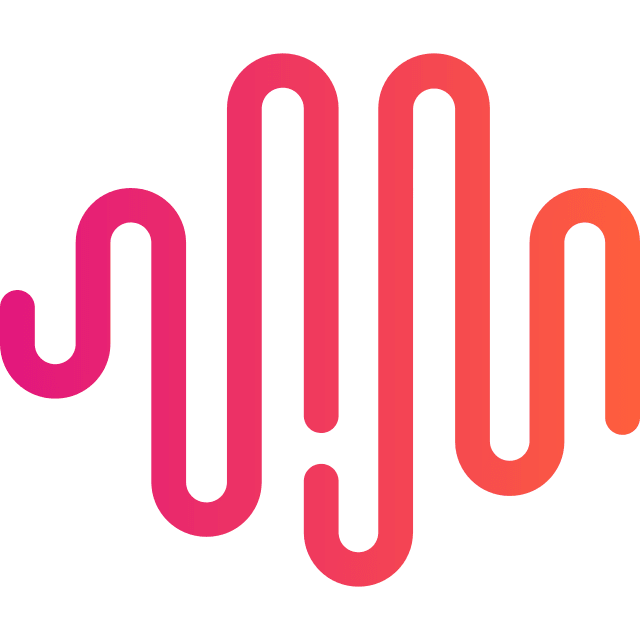}~MiniMax-M3 & VL & 31.12 & 18.83 & 33.96 & 50.87 & 29.49 & 21.25 & 37.38 & 29.32 \\
    \modelicon{1.0em}{qwen-color.png}~Qwen3.8-Max & VL & 43.51 & 27.59 & 46.52 & 55.77 & 38.60 & 34.84 & 41.90 & 38.37 \\
    \rowcolor{tblZebra}\modelicon{1.0em}{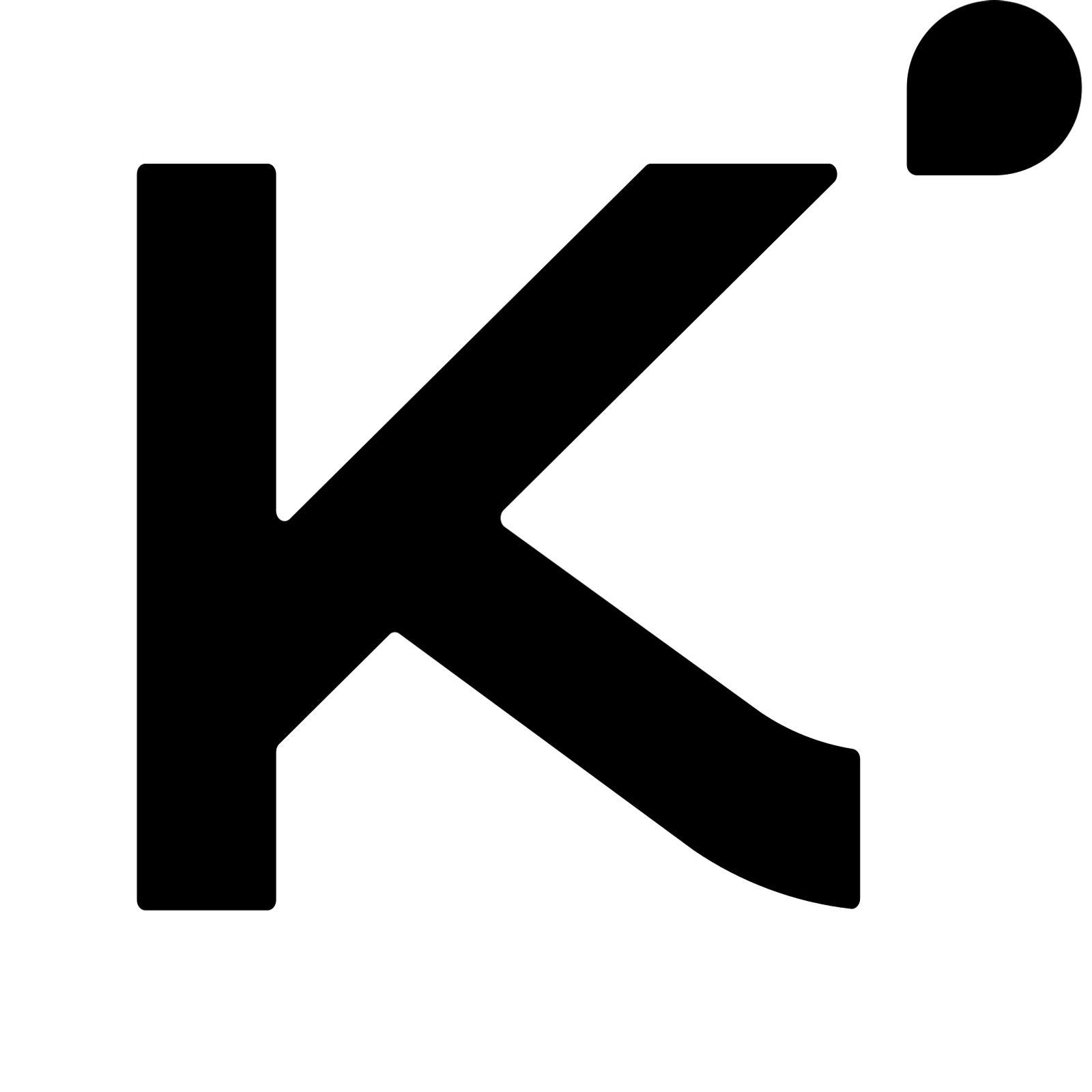}~Kimi-K3 & VL & 43.32 & 30.26 & 49.23 & 56.59 & 40.63 & 38.64 & 45.25 & 41.95 \\
    \midrule
    \rowcolor{tblGroup}\multicolumn{10}{c}{\tblHead{Closed-Source Models}} \\
    \modelicon{1.0em}{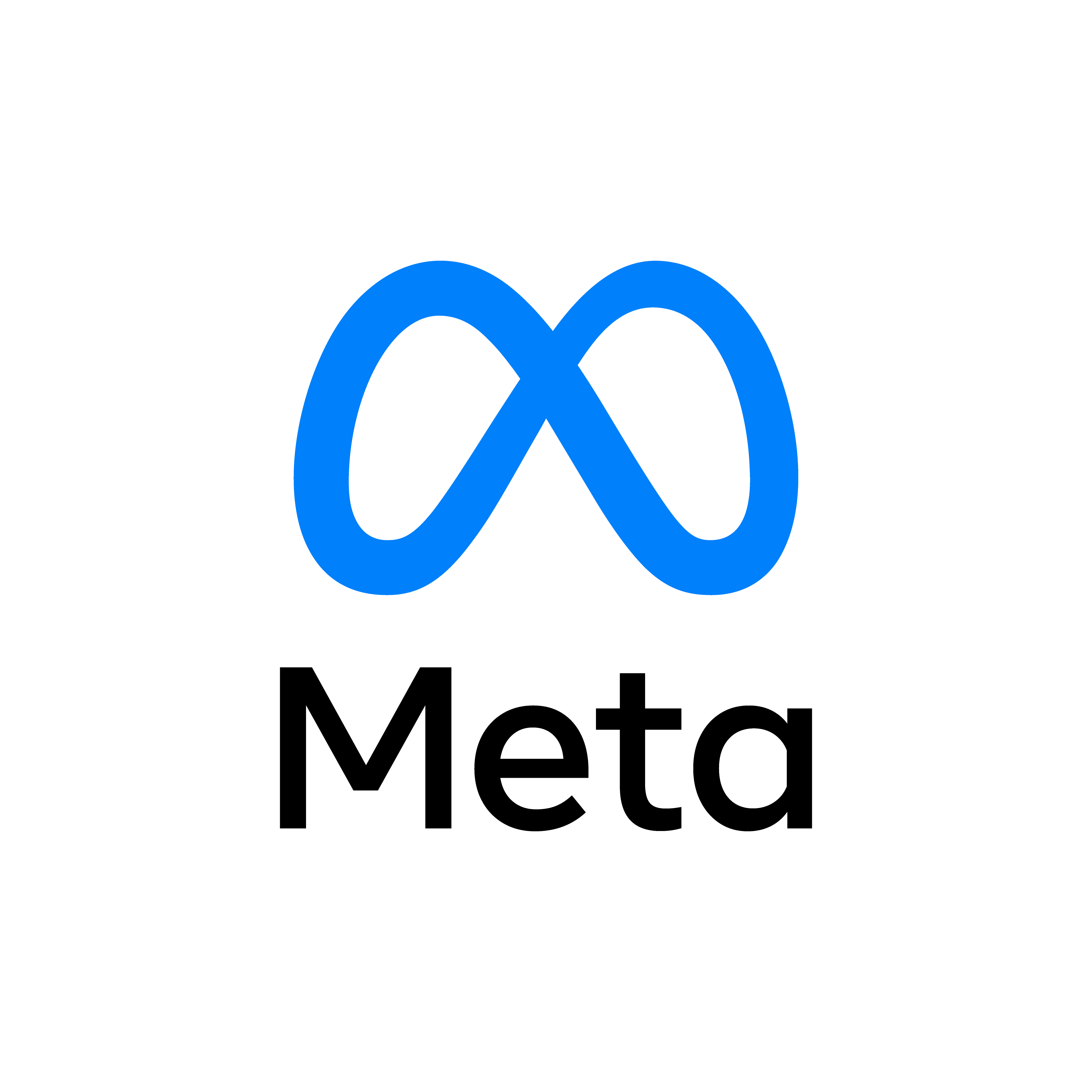}~Muse Spark 1.1 & Omni & 22.29 & 10.14 & 45.99 & 37.91 & 24.08 & 18.45 & 32.37 & 25.41 \\
    \rowcolor{tblZebra}\modelicon{1.0em}{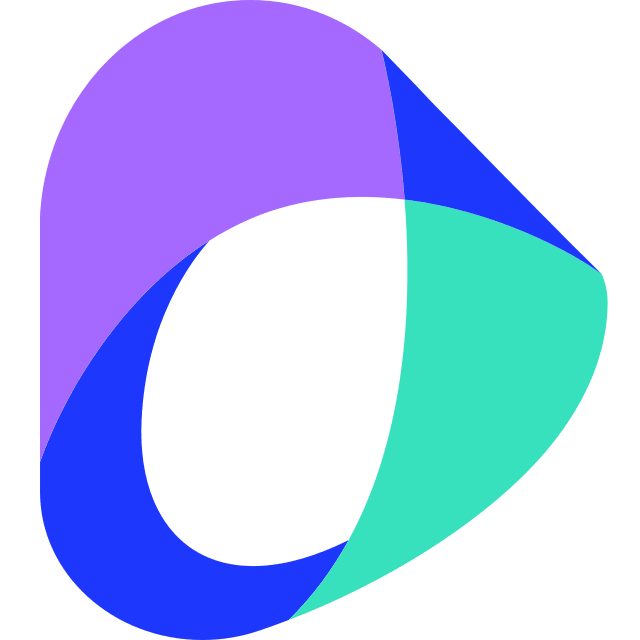}~Seed2.0 Lite & Omni & 20.38 & 11.70 & 63.36 & 38.87 & 28.33 & 47.91 & 45.14 & 46.52 \\
    \modelicon{1.0em}{gemini-color.png}~Gemini 3.5 Flash & Omni & \textbf{59.20} & \textbf{40.52} & 53.71 & \textbf{71.26} & \textbf{51.17} & 47.04 & 47.88 & 47.46 \\
    \rowcolor{tblZebra}\modelicon{1.0em}{qwen-color.png}~Qwen3.5-Omni-Plus & Omni & 20.79 & 11.86 & 75.46 & 32.28 & 29.56 & \textbf{49.99} & 48.29 & \textbf{49.14} \\
    \modelicon{1.0em}{doubao-color.png}~Seed2.1 Pro & VL & 37.54 & 23.67 & 37.40 & 57.03 & 34.48 & 33.21 & 44.96 & 39.08 \\
    \rowcolor{tblZebra}\modelicon{1.0em}{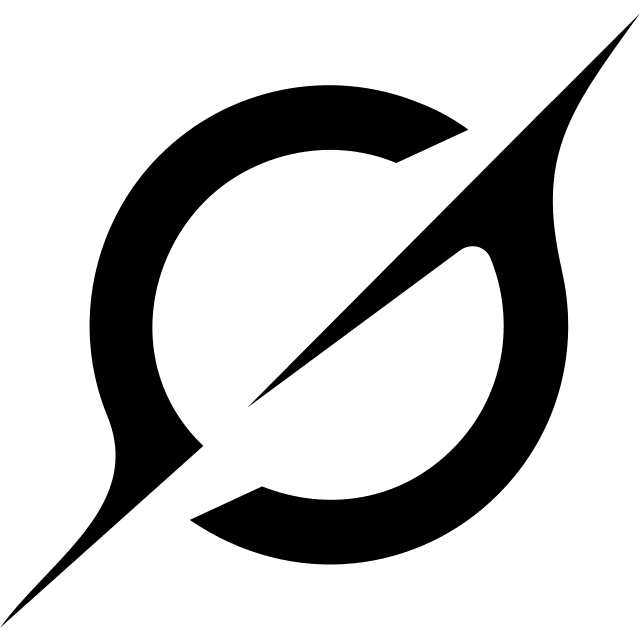}~Grok 4.6 & VL & 42.52 & 29.92 & 59.78 & 51.43 & 41.45 & 34.60 & 47.06 & 40.83 \\
    \modelicon{1.0em}{gemini-color.png}~Gemini 3.1 Pro & Omni & 26.93 & 16.46 & 74.09 & 41.67 & 34.08 & 42.19 & 44.18 & 43.19 \\
    \rowcolor{tblZebra}\modelicon{1.0em}{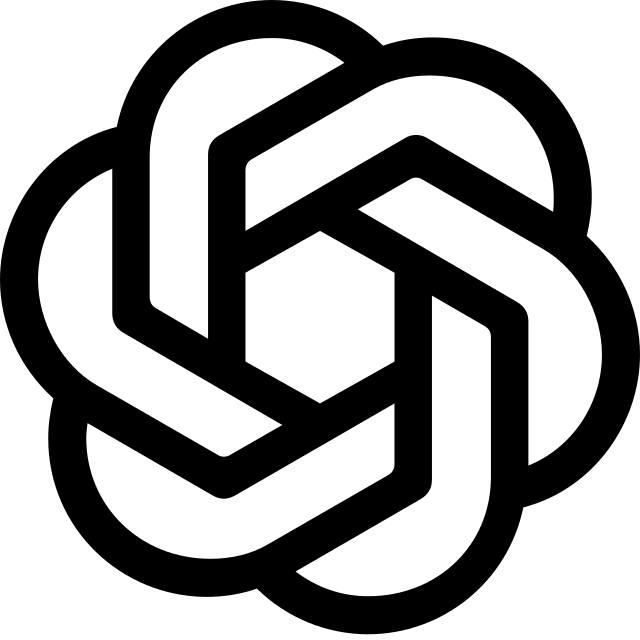}~GPT-5.6 Sol & VL & 55.92 & 37.32 & 29.55 & 63.49 & 42.86 & 47.51 & 49.17 & 48.34 \\
    \modelicon{1.0em}{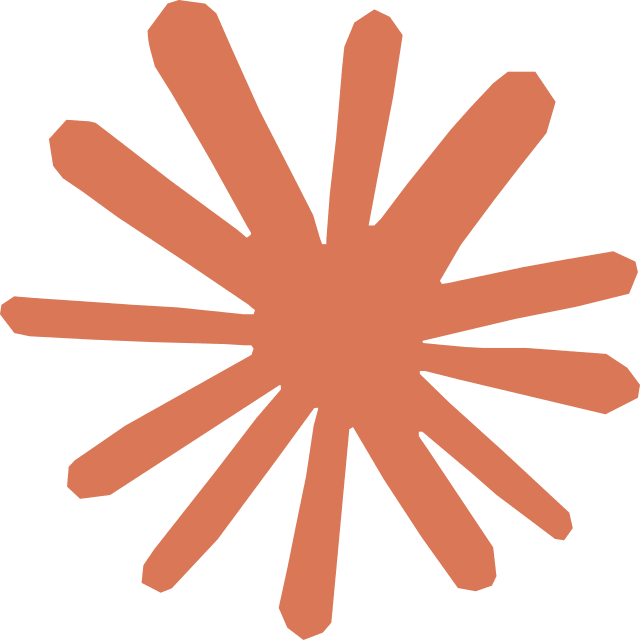}~Claude Opus 5 & VL & 54.11 & 38.90 & 50.88 & 61.67 & 47.37 & 48.06 & \textbf{49.76} & 48.91 \\
    \bottomrule
  \end{tabularx}
  \arrayrulecolor{black}
\end{table}

% Generated by scripts/sync_main_c_table.py; do not edit score cells manually.
\begin{table}[t]
  \centering
  \scriptsize
  \setlength{\tabcolsep}{2.2pt}
  \renewcommand{\arraystretch}{1.08}
  \caption{Instruction Utility (Track C) with the fixed DeepSeek-v4-pro
  executor. EFS and its components are absolute scores, whereas NIU is
  normalized by oracle EFS.}
  \label{tab:main-c-results}
  \arrayrulecolor{tblNavy}
  \begin{tabularx}{\textwidth}{@{}>{\raggedright\arraybackslash}p{2.45cm}c
      *{4}{Y}!{\hspace{2pt}{\color{tblNavy}\vrule width 0.3pt}\hspace{2pt}}*{2}{Y}@{}}
    \toprule
    & & \multicolumn{4}{>{\columncolor{tblZebra}}c}{\tblHead{Absolute EFS Components}}
      & \multicolumn{2}{>{\columncolor{tblCTint}}c}{\textcolor{tblTrackC}{\textbf{Track C: Instruction Utility}}} \\
    \cmidrule(lr){3-6}\cmidrule(lr){7-8}
    \tblHead{Model} & \tblHead{Modality}
      & \tblHead{Grd.} & \tblHead{Edit} & \tblHead{No dmg.}
      & \tblHead{Rev.} & \tblHead{EFS} & \tblHead{NIU} \\
    \midrule
    \rowcolor{tblGroup}\multicolumn{8}{c}{\tblHead{Oracle Executor}} \\
    \modelicon{1.0em}{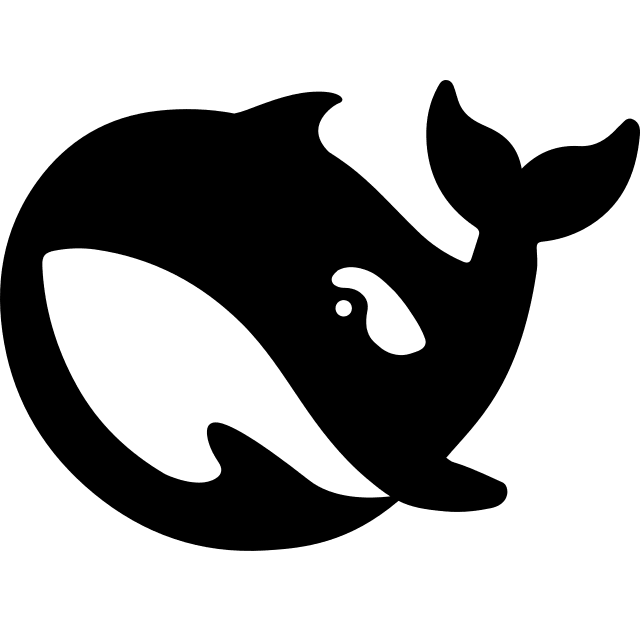}~DeepSeek-v4-pro & Text & \textbf{95.79} & \textbf{91.69} & \textbf{74.01} & \textbf{97.35} & \textbf{89.69} & -- \\
    \midrule
    \rowcolor{tblGroup}\multicolumn{8}{c}{\tblHead{Open-Source Models}} \\
    \modelicon{1.0em}{minicpm.png}~MiniCPM-o 4.5 & Omni & 38.20 & 21.21 & 41.37 & 48.75 & 32.45 & 36.18 \\
    \rowcolor{tblZebra}\modelicon{1.0em}{gemini-color.png}~Gemma 4 12B & Omni & 2.58 & 0.71 & 61.89 & 15.60 & 16.11 & 17.96 \\
    \modelicon{1.0em}{qwen-color.png}~Qwen3.8-27B & VL & 28.93 & 15.65 & 43.26 & 45.55 & 28.48 & 31.75 \\
    \rowcolor{tblZebra}\modelicon{1.0em}{nvidia.png}~Nemotron 3 Nano & Omni & 2.65 & 0.77 & \textbf{69.30} & 15.38 & 17.59 & 19.61 \\
    \modelicon{1.0em}{xiaomimimo.png}~MiMo-V2.5 & Omni & 46.42 & 28.16 & 43.04 & 52.69 & 37.87 & 42.22 \\
    \rowcolor{tblZebra}\modelicon{1.0em}{minimax-color.png}~MiniMax-M3 & VL & 31.86 & 16.79 & 39.36 & 46.10 & 28.67 & 31.97 \\
    \modelicon{1.0em}{qwen-color.png}~Qwen3.8-Max & VL & 46.05 & 28.04 & 45.67 & 58.81 & 39.52 & 44.06 \\
    \rowcolor{tblZebra}\modelicon{1.0em}{kimi.png}~Kimi-K3 & VL & 47.59 & 30.36 & 44.61 & 60.26 & 40.91 & 45.62 \\
    \midrule
    \rowcolor{tblGroup}\multicolumn{8}{c}{\tblHead{Closed-Source Models}} \\
    \modelicon{1.0em}{meta.png}~Muse Spark 1.1 & Omni & 33.71 & 12.72 & 29.30 & 43.88 & 24.37 & 27.17 \\
    \rowcolor{tblZebra}\modelicon{1.0em}{doubao-color.png}~Seed2.0 Lite & Omni & 55.30 & 37.68 & 52.10 & 63.18 & 47.42 & 52.87 \\
    \modelicon{1.0em}{gemini-color.png}~Gemini 3.5 Flash & Omni & 59.16 & 39.58 & 51.42 & \textbf{68.92} & 49.77 & 55.49 \\
    \rowcolor{tblZebra}\modelicon{1.0em}{qwen-color.png}~Qwen3.5-Omni-Plus & Omni & \textbf{61.39} & \textbf{41.72} & 53.06 & 67.36 & \textbf{51.08} & \textbf{56.95} \\
    \modelicon{1.0em}{doubao-color.png}~Seed2.1 Pro & VL & 43.94 & 28.39 & 43.66 & 56.68 & 38.66 & 43.10 \\
    \rowcolor{tblZebra}\modelicon{1.0em}{grok.png}~Grok 4.6 & VL & 44.72 & 28.12 & 45.53 & 54.81 & 38.60 & 43.04 \\
    \modelicon{1.0em}{gemini-color.png}~Gemini 3.1 Pro & Omni & 53.81 & 34.84 & 47.91 & 65.74 & 45.53 & 50.76 \\
    \rowcolor{tblZebra}\modelicon{1.0em}{openai.png}~GPT-5.6 Sol & VL & 56.70 & 37.94 & 45.98 & 61.78 & 46.19 & 51.50 \\
    \modelicon{1.0em}{claude-color.png}~Claude Opus 5 & VL & 56.78 & 39.34 & 46.13 & 62.49 & 47.07 & 52.48 \\
    \bottomrule
  \end{tabularx}
  \arrayrulecolor{black}
\end{table}

\subsection{Direct Editing Performance}
\label{sec:direct-editing-performance}

\textbf{Correct grounding often fails to produce a correct edit.} Gemini 3.5 Flash ranks first in Direct Editing. Yet only 42.59\% of its 11,382 scored steps receive both grounding and edit credit; in another 16.68\%, it finds the correct target but fails to execute the requested edit (Figure~\ref{fig:track-ac-diagnostics}(a)). Edit credit without grounding credit never exceeds 3.83\% for any model. Thus, execution after localization is the main bottleneck, as models often identify the target without implementing the requested change.

\textbf{High No-damage can mask inaction.} Qwen3.8-27B has the highest No-damage score (82.94) but only 7.56 in Edit Fulfillment; Qwen3.5-Omni-Plus scores 75.46 and 11.86, while Nemotron 3 Nano scores 71.27 and 0.04. These models leave much of the original page intact partly because they perform few requested edits. EFS therefore evaluates preservation jointly with grounding, edit fulfillment, and revision handling instead of rewarding No-damage in isolation.

\begin{figure*}[t]
  \centering
  \includegraphics[width=\textwidth]{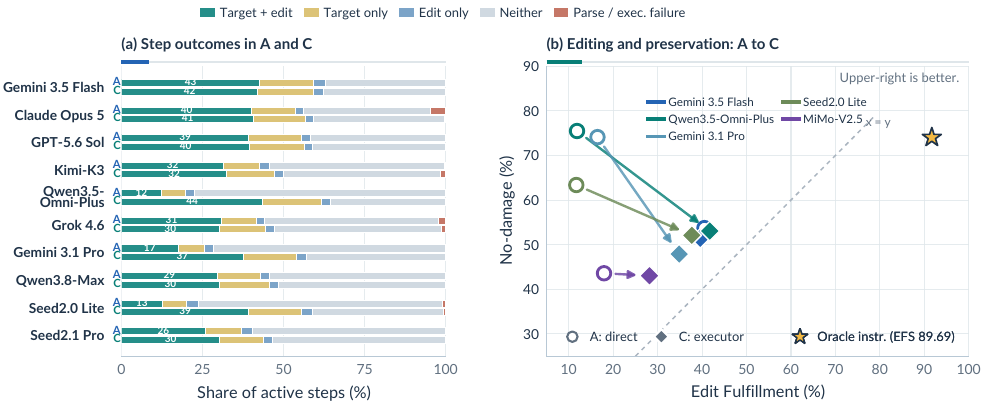}
  \vspace{-0.7em}
  \caption{Track~A/C diagnostics. (a) Step outcomes for the ten models with highest summed A/C EFS. (b) A-to-C changes for the five highest-scoring Omni models by summed A/C EFS. Open circles and filled diamonds denote Tracks~A and C; the star denotes oracle instructions.}
  \label{fig:track-ac-diagnostics}
\end{figure*}

\subsection{Instruction Recovery and Downstream Utility}
\label{sec:instruction-recovery-utility}

\textbf{Some Omni models recover intent better than they execute it.} Among 15 complete models, Track~A EFS and Track~B IRS correlate moderately ($\rho=0.686$), with Qwen3.5-Omni-Plus and Seed2.0 Lite rising from ninth and twelfth to first and fifth, respectively (Figure~\ref{fig:cross-track-diagnostic}(a)). The fixed executor raises their EFS by 21.52 and 19.09 points. Track~C exceeds Track~A for 11 of 15 pairs and improves joint correctness by 3.77--14.77 points across all 12 edit types (Appendix~\ref{app:operation-analysis}).

\begin{figure*}[t]
  \centering
  \includegraphics[width=\textwidth]{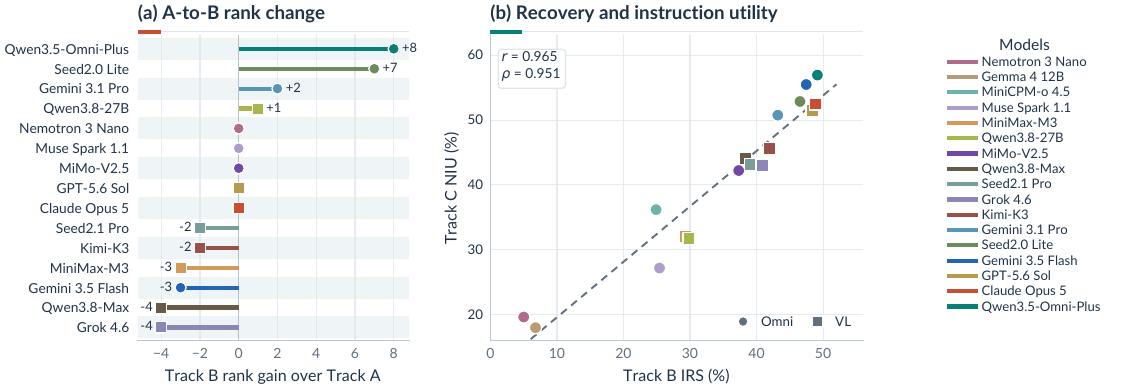}
  \vspace{-0.7em}
  \caption{Cross-track relationships. (a) A-to-B rank change (positive favors B). (b) Track~B IRS versus Track~C NIU with a least-squares fit. Colors denote models; shapes denote Omni/VL input.}
  \label{fig:cross-track-diagnostic}
\end{figure*}

\textbf{Lower No-damage is not necessarily a regression.} Figure~\ref{fig:track-ac-diagnostics}(b) shows that Qwen3.5-Omni-Plus increases Edit Fulfillment from 11.86 to 41.72 and EFS from 29.56 to 51.08, while No-damage decreases from 75.46 to 53.06. Seed2.0 Lite follows the same pattern: Edit Fulfillment rises from 11.70 to 37.68 and EFS from 28.33 to 47.42, while No-damage falls from 63.36 to 52.10. Both cases show more requested editing and higher EFS despite lower preservation.

\textbf{Recovery predicts downstream utility, but a large oracle gap remains.} Across 17 models, Track~B IRS strongly correlates with Track~C NIU (Pearson $r=0.965$; Spearman $\rho=0.951$; Figure~\ref{fig:cross-track-diagnostic}(b)). Qwen3.5-Omni-Plus scores 51.08 EFS and 56.95 NIU on Track~C versus 89.69 oracle EFS with the same executor. On a 300-instance subset, self-execution leaves EFS gaps between oracle and recovered instructions of 36.96 for Gemini and 22.24 for Qwen. Thus, the external executor alone does not explain the gap (Appendix~\ref{app:self-execution-control}).

\section{Analysis and Robustness}
\label{sec:analysis-robustness}

\subsection{Visual Dependence and Audiovisual Ablations}
\label{sec:performance-by-visual-dependence}
\label{sec:controlled-modality-ablation}

\textbf{Omni models show their strongest advantage on Hard instances.} On the 481 Hard instances, Gemini 3.5 Flash has the highest observed Track~A score at 39.74 EFS, while Qwen3.5-Omni-Plus has the highest Track~B score at 37.36 IRS (Figure~\ref{fig:difficulty-dependence} and Figure~\ref{fig:difficulty-diagnostics}). Proximity to $y=x$ in Figure~\ref{fig:difficulty-dependence}(b,d) indicates less Hard-subset degradation. Gemini 3.5 Flash drops by 11.43 points on Track~A, versus 14.92--15.87 for the three leading VL models; on Track~B, Qwen3.5-Omni-Plus drops by 11.78 versus Claude Opus 5's 14.27, increasing its observed margin from 0.23 overall to 2.72 on Hard. This supports the use of cursor--speech timing beyond text-recoverable cues.

\begin{figure}[t]
  \centering
  \includegraphics[width=\textwidth]{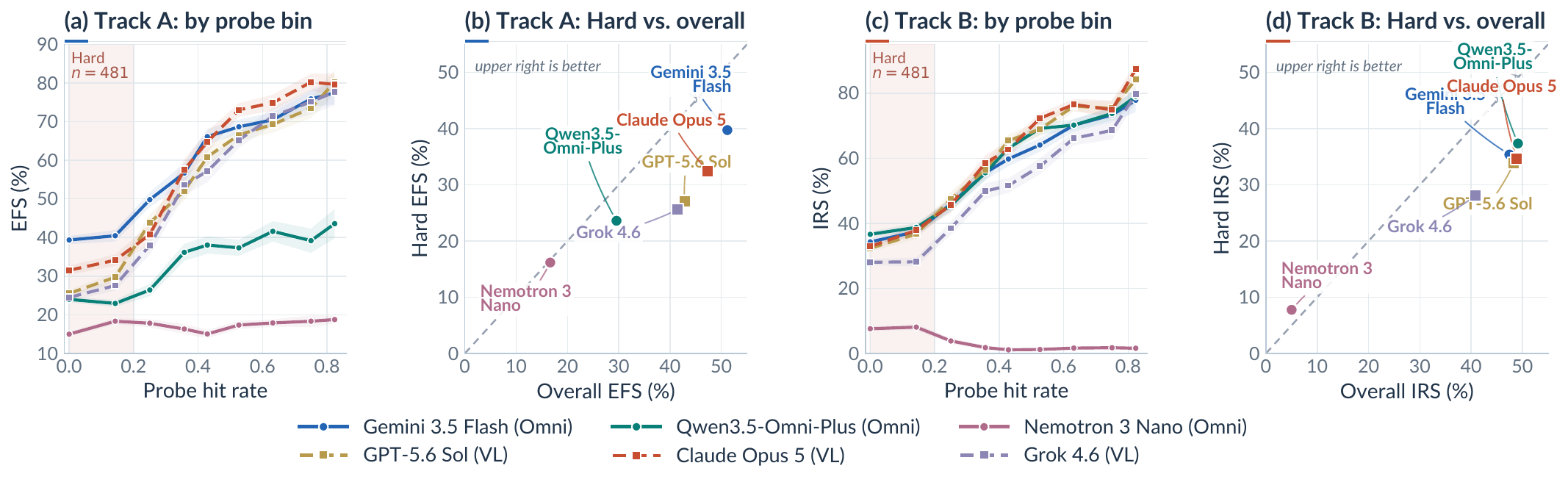}
  \vspace{-0.7em}
  \caption{Scores by probe bin (a,c) and overall versus Hard (b,d; $h_i<0.2$, $n=481$).}
  \label{fig:difficulty-dependence}
\end{figure}

\textbf{Synchronized audiovisual input is strongest for both tested Omni models.} Relative to transcript only, native audiovisual input improves Track~A, B, and C by 18.41, 29.28, and 31.78 points for Gemini 3.5 Flash and by 5.97, 20.49, and 20.01 points for Qwen3.5-Omni-Plus. Relative to silent video with transcript, the gains span 9.39--16.80 and 7.13--19.82 points, respectively (Table~\ref{tab:input-modality-ablation}). A complementary timing intervention separates synchronization from content: shifting audio by $\pm5$ seconds while preserving its content lowers IRS by 11.65 points for Gemini and 13.73 for Qwen across 300 instances. Conversely, adding timestamps to the same manual transcripts and 1 FPS frames raises IRS by 8.76 points for GPT-5.6 Sol and 14.98 for Grok 4.6 (Table~\ref{tab:timestamped-transcript-ablation}). Both timing interventions affect target grounding more than action recovery, supporting the importance of temporal correspondence for resolving references (Appendix~\ref{app:modality-ablation}).

\begin{table}[t]
  \centering
  \caption{Input-modality ablation on the 918 instances. Tracks~A/C report EFS and Track~B reports Component-F1 IRS; bold marks each model's best input.}
  \label{tab:input-modality-ablation}
  \small
  \setlength{\tabcolsep}{7pt}
  \arrayrulecolor{tblNavy}
  \begin{tabular}{llccc}
    \toprule
    \tblHead{Model} & \tblHead{Input} & \tblTrackHead{A}{Track A} & \tblTrackHead{B}{Track B} & \tblTrackHead{C}{Track C} \\
    \midrule
    \modelicon{1.0em}{gemini-color.png}~Gemini 3.5 Flash & Transcript only & 32.76 & 18.18 & 18.00 \\
    \rowcolor{tblZebra} & Silent video + transcript & 34.37 & 38.06 & 34.96 \\
     & \textbf{Audio-video} & \textbf{51.17} & \textbf{47.46} & \textbf{49.77} \\
    \midrule
    \modelicon{1.0em}{qwen-color.png}~Qwen3.5-Omni-Plus & Transcript only & 23.59 & 28.65 & 31.07 \\
    \rowcolor{tblZebra} & Silent video + transcript & 22.43 & 32.58 & 31.26 \\
     & \textbf{Audio-video} & \textbf{29.56} & \textbf{49.14} & \textbf{51.08} \\
    \bottomrule
  \end{tabular}
  \arrayrulecolor{black}
\end{table}

\subsection{Judge Robustness}
\label{sec:judge-robustness}

\textbf{Judge replacement preserves the leader and broad model ordering.} To test self-preference~\citep{panickssery2024llm}, we use a fixed evaluator-validation set of 300 instances for six models. Relative to Gemini 3.1 Pro, GPT-5.5 yields Spearman $\rho=0.943$ and Kendall $\tau=0.867$, while DeepSeek-v4-pro preserves the ordering exactly; Gemini 3.5 Flash remains first (Table~\ref{tab:judge-robustness}).

Absolute scores vary more: mean absolute differences from the primary judge are 5.16 EFS for GPT-5.5 and 2.09 for DeepSeek-v4-pro. We find no consistent same-family advantage; relative to the other judges' mean, same-family changes are $+1.43$ for Gemini 3.5 Flash, $-4.68$ for Gemini 3.1 Pro, and $-0.31$ for GPT-5.6 Sol. Appendix~\ref{app:judge-ablation} reports the complete matrix. A blinded human evaluation of 1,451 steps finds strong output-level correlations for Track~A grounding and edit fulfillment and Track~B IRS (up to $r=0.906$; Appendix~\ref{app:human-judge-calibration}).

% Source SHA256: f6e0e58e0980575abe37d6f614d8a8e3dfffb5f8aa22a3516327c174170deda6
\begin{table}[t]
  \centering
  \caption{Judge robustness for six models on 300 Track~A instances. Correlations use Gemini 3.1 Pro as reference; $\lvert\Delta\rvert$ is in EFS points.}
  \label{tab:judge-robustness}
  \small
  \setlength{\tabcolsep}{5.5pt}
  \arrayrulecolor{tblNavy}
  \begin{tabular}{lccccc}
    \toprule
    \tblHead{Alternative judge} & \tblHead{Pearson $r$} & \tblHead{Spearman $\rho$} & \tblHead{Kendall $\tau$} & \tblHead{Top model} & \tblHead{Mean $\lvert\Delta\rvert$} \\
    \midrule
    \modelicon{1.0em}{openai.png}~GPT-5.5 & 0.983 & 0.943 & 0.867 & Gemini 3.5 Flash & 5.16 \\
    \rowcolor{tblZebra}\modelicon{1.0em}{deepseek.png}~DeepSeek-v4-pro & 0.998 & 1.000 & 1.000 & Gemini 3.5 Flash & 2.09 \\
    \bottomrule
  \end{tabular}
  \arrayrulecolor{black}
\end{table}

\subsection{Metric and Aggregation Robustness}
\label{sec:metric-aggregation-robustness}

\textbf{The Track~A and C leaders are stable across tested EFS parameterizations.} Across 156 weight and No-damage settings, the leaders never change; 150 retain Spearman $\rho\geq0.9$ on Track~A, and all do so on Track~C. The largest shift, under a No-damage-heavy vector with $\alpha=0.05$, moves Qwen3.8-27B from thirteenth to fifth on Track~A, so middle ranks should not be overinterpreted. Alternative Track~B formulations yield $\rho\geq0.983$ but change the leader under precision-heavy or joint scoring (Appendices~\ref{app:efs-sensitivity} and~\ref{app:track-b-sensitivity}).

\textbf{Page dependence and zero-filled outputs mainly affect close races.} Page bootstrap ranks Gemini first on A in 99.99\% of resamples and Qwen first on B and C in 55.67\% and 98.63\%, respectively. Qwen and Seed show positive A-to-C gains throughout. Valid-only scores retain all leaders and Track~B's top three; no model--track exceeds 41 zero-filled outputs (Appendices~\ref{app:source-page-bootstrap} and~\ref{app:coverage-robustness}).

\section{Conclusion}
\label{sec:conclusion}

Omni2Web evaluates weakly referential screen-recorded web editing through direct editing, instruction recovery, and downstream utility. Results across 17 models reveal substantial headroom in both multimodal understanding and execution: models often locate the correct element yet fail to implement the requested change. Controlled modality and temporal-alignment experiments further show that synchronized speech and visual context are important for recovering user intent. Meanwhile, some Omni models produce instructions that a coding model executes substantially better than their direct edits, demonstrating the promise of separating intent recovery from code generation. The remaining gap to oracle instructions leaves considerable room to improve both stages. Omni2Web shifts web-editing evaluation from complete specifications to incomplete, situated user intent.

\endgroup
\bibliography{references}
\bibliographystyle{preprint}

\newpage
\appendix
\section{Appendix}
\label{app:appendix}

\setlength{\abovecaptionskip}{10pt}
\setlength{\belowcaptionskip}{5pt}

\hypersetup{hidelinks, linkcolor=black, citecolor=black, urlcolor=black}
\begingroup
\small
\setlength{\parskip}{1pt}
\noindent\textbf{Contents}\par
\noindent\hyperref[app:discussion-threats]{\ref*{app:discussion-threats}\quad Discussion and Threats to Validity}\dotfill\pageref{app:discussion-threats}\par
\noindent\hyperref[app:operation-analysis]{\ref*{app:operation-analysis}\quad Operation-Level Analysis}\dotfill\pageref{app:operation-analysis}\par
\noindent\hyperref[app:modality-ablation]{\ref*{app:modality-ablation}\quad Controlled Modality and Temporal Alignment Ablations}\dotfill\pageref{app:modality-ablation}\par
\noindent\hyperref[app:self-execution-control]{\ref*{app:self-execution-control}\quad Same-Model Execution Control}\dotfill\pageref{app:self-execution-control}\par
\noindent\hyperref[app:judge-ablation]{\ref*{app:judge-ablation}\quad Alternative-Judge Ablation}\dotfill\pageref{app:judge-ablation}\par
\noindent\hyperref[app:human-judge-calibration]{\ref*{app:human-judge-calibration}\quad Human Calibration of the LLM Judge}\dotfill\pageref{app:human-judge-calibration}\par
\noindent\hyperref[app:efs-sensitivity]{\ref*{app:efs-sensitivity}\quad EFS Hyperparameter Sensitivity}\dotfill\pageref{app:efs-sensitivity}\par
\noindent\hyperref[app:track-b-sensitivity]{\ref*{app:track-b-sensitivity}\quad Track B Metric Sensitivity}\dotfill\pageref{app:track-b-sensitivity}\par
\noindent\hyperref[app:source-page-bootstrap]{\ref*{app:source-page-bootstrap}\quad Source-Page Cluster Bootstrap}\dotfill\pageref{app:source-page-bootstrap}\par
\noindent\hyperref[app:coverage-robustness]{\ref*{app:coverage-robustness}\quad Coverage and Zero-Fill Robustness}\dotfill\pageref{app:coverage-robustness}\par
\noindent\hyperref[app:prompts]{\ref*{app:prompts}\quad Prompt Documentation}\dotfill\pageref{app:prompts}\par
\noindent\hspace*{1em}\hyperref[app:contestant-prompts]{\ref*{app:contestant-prompts}\quad Contestant and Executor Prompts}\dotfill\pageref{app:contestant-prompts}\par
\noindent\hspace*{2em}\hyperref[app:track-a-prompt]{Track A: Direct Editing}\dotfill\pageref{app:track-a-prompt}\par
\noindent\hspace*{2em}\hyperref[app:track-b-prompt]{Track B: Instruction Recovery}\dotfill\pageref{app:track-b-prompt}\par
\noindent\hspace*{2em}\hyperref[app:track-c-prompt]{Track C: Fixed Executor}\dotfill\pageref{app:track-c-prompt}\par
\noindent\hspace*{1em}\hyperref[app:scoring-prompts]{\ref*{app:scoring-prompts}\quad Scoring Prompts}\dotfill\pageref{app:scoring-prompts}\par
\noindent\hspace*{2em}\hyperref[app:rubric-judge-prompt]{Tracks A and C Rubric Judge}\dotfill\pageref{app:rubric-judge-prompt}\par
\noindent\hspace*{2em}\hyperref[app:matching-judge-prompt]{Track B Matching Judge}\dotfill\pageref{app:matching-judge-prompt}\par
\noindent\hspace*{2em}\hyperref[app:no-damage-prompt]{No-Damage Judge}\dotfill\pageref{app:no-damage-prompt}\par
\noindent\hspace*{2em}\hyperref[app:image-judge-prompt]{Replacement-Image Content Judge}\dotfill\pageref{app:image-judge-prompt}\par
\noindent\hspace*{1em}\hyperref[app:difficulty-prompts]{\ref*{app:difficulty-prompts}\quad Difficulty-Probe Prompts}\dotfill\pageref{app:difficulty-prompts}\par
\noindent\hspace*{2em}\hyperref[app:target-recovery-prompt]{Text-Only Target Recovery}\dotfill\pageref{app:target-recovery-prompt}\par
\noindent\hspace*{2em}\hyperref[app:target-equivalence-prompt]{Target-Equivalence Judge}\dotfill\pageref{app:target-equivalence-prompt}\par
\endgroup

\subsection{Discussion and Threats to Validity}
\label{app:discussion-threats}

\textbf{The three-track design provides complementary evidence about intent recovery and execution.} Track~A evaluates direct webpage editing from a screen recording, Track~B isolates explicit-instruction recovery, and Track~C tests whether a fixed code executor can turn those instructions into correct modifications. Some Omni models have limited direct-editing ability yet recover user intent useful to the downstream executor. Combining multimodal intent understanding with specialized code execution therefore offers a practical route for handling weakly referential editing requests. The tracks are complementary diagnostic views, not interchangeable leaderboards.

\textbf{Model-based scoring only approximates editing quality.} EFS and IRS convert complex webpage edits into reproducible, rubric-based scores, but cannot fully replace human evaluation~\citep{zheng2023judging}. We use Gemini 3.1 Pro~\citep{google2026gemini31pro} as the primary judge and test robustness with GPT-5.5~\citep{openai2026gpt55} and DeepSeek-v4-pro~\citep{deepseek2026deepseek}. Absolute scores differ, but the leading model and broad ordering remain stable. The EFS category weights and the $0.15$ damage penalty are prespecified heuristics rather than data-derived quantities. Appendix~\ref{app:efs-sensitivity} varies both choices across 156 configurations. Appendix~\ref{app:track-b-sensitivity} likewise finds high rank correlations across alternative IRS formulations, although some alternatives change the exact leader.

\textbf{Track~C depends on the fixed executor.} All Track~C results execute recovered instructions with DeepSeek-v4-pro~\citep{deepseek2026deepseek}, holding execution capability fixed across models. NIU, absolute Track~C scores, and rankings may nevertheless change with the executor. The 89.69 EFS from oracle instructions is an executor-specific reference, not a theoretical task ceiling.

\textbf{The benchmark's scope limits external validity.} Omni2Web contains 918 instances constructed from 129 source webpages and covers 12 webpage-editing operation types. It focuses on multi-step modifications to existing HTML pages and does not represent the full range of open-ended software engineering, live website interaction, or general GUI-agent tasks. Online API results are snapshots of particular model versions and evaluation dates. Zero-filling refused, empty, or otherwise unusable outputs, together with valid-only robustness and coverage reporting, reduces selective bias but cannot eliminate model-version drift or output nondeterminism.

\subsection{Operation-Level Analysis}
\label{app:operation-analysis}

We align each of the 13,907 reference steps with one of the 12 operation types and compare the 15 models that have results on all three tracks. For Tracks~A and C, a step is jointly correct when both its target-grounding and edit-fulfillment or revision decisions pass; for Track~B, both the recovered target and action must match. We first compute micro accuracy over steps of each operation within a model and then macro-average across models. These diagnostic rates are not EFS or IRS: they exclude the instance-level No-damage term and, unlike Component-F1 IRS, have no prediction-level precision term because extra predictions cannot be assigned to a reference operation. Table~\ref{tab:operation-level-summary} reports the cross-model macro-average, while Table~\ref{tab:operation-level-selected-models} gives the corresponding per-model rates for the two leading Omni systems.

\begin{table}[H]
  \centering
  \scriptsize
  \setlength{\tabcolsep}{6.0pt}
  \renewcommand{\arraystretch}{1.06}
  \caption{Macro-averaged step-level joint correctness (\%) by operation over the 15 models evaluated on all three tracks. B retains the full history; A/C use active steps. $\Delta$ is Track~C minus Track~A.}
  \label{tab:operation-level-summary}
  \arrayrulecolor{tblNavy}
  \begin{tabularx}{\textwidth}{@{}>{\raggedright\arraybackslash}X*{6}{Y}@{}}
    \toprule
    \tblHead{Operation} & \tblHead{B steps} & \tblHead{A/C steps} & \tblTrackHead{A}{Track A} & \tblTrackHead{B}{Track B} & \tblTrackHead{C}{Track C} & \tblHead{$\Delta$ C--A} \\
    \midrule
    Batch style & 1,080 & 909 & 17.32 & 19.99 & 21.09 & +3.77 \\
    \rowcolor{tblZebra} Change color & 645 & 643 & 24.99 & 35.00 & 34.18 & +9.20 \\
    Change font & 38 & 37 & 27.39 & 42.11 & 42.16 & +14.77 \\
    \rowcolor{tblZebra} Change layout & 1,985 & 1,553 & 25.43 & 41.43 & 33.05 & +7.61 \\
    Change theme & 1,178 & 884 & 19.46 & 28.30 & 26.13 & +6.67 \\
    \rowcolor{tblZebra} Hide/delete & 1,135 & 996 & 19.59 & 22.65 & 25.41 & +5.82 \\
    Move & 1,423 & 958 & 11.89 & 16.03 & 16.88 & +4.99 \\
    \rowcolor{tblZebra} Region edit & 1,325 & 983 & 19.92 & 28.36 & 27.60 & +7.68 \\
    Replace image & 1,040 & 890 & 22.51 & 47.26 & 29.42 & +6.91 \\
    \rowcolor{tblZebra} Revert & 2,165 & 2,164 & 31.55 & 28.79 & 40.56 & +9.01 \\
    Rewrite text & 287 & 287 & 36.77 & 42.53 & 42.21 & +5.44 \\
    \rowcolor{tblZebra} Swap & 1,606 & 1,078 & 19.06 & 32.03 & 26.80 & +7.74 \\
    \bottomrule
  \end{tabularx}
  \arrayrulecolor{black}
\end{table}

\begin{table}[H]
  \centering
  \scriptsize
  \setlength{\tabcolsep}{3.0pt}
  \renewcommand{\arraystretch}{1.06}
  \caption{Per-model step-level joint correctness (\%) by operation for Gemini 3.5 Flash and Qwen3.5-Omni-Plus.}
  \label{tab:operation-level-selected-models}
  \arrayrulecolor{tblNavy}
  \begin{tabularx}{\textwidth}{@{}>{\raggedright\arraybackslash}p{1.65cm}*{6}{Y}@{}}
    \toprule
    \tblHead{Operation} & \multicolumn{3}{>{\columncolor{tblZebra}}c}{\modelicon{1.0em}{gemini-color.png}~\tblHead{Gemini 3.5 Flash}} & \multicolumn{3}{>{\columncolor{tblZebra}}c}{\modelicon{1.0em}{qwen-color.png}~\tblHead{Qwen3.5-Omni-Plus}} \\
    & \tblTrackHead{A}{A} & \tblTrackHead{B}{B} & \tblTrackHead{C}{C} & \tblTrackHead{A}{A} & \tblTrackHead{B}{B} & \tblTrackHead{C}{C} \\
    \midrule
    Batch style & 33.33 & 26.85 & 27.50 & 6.60 & 28.98 & 29.15 \\
    \rowcolor{tblZebra} Change color & 57.08 & 52.71 & 52.41 & 18.20 & 60.62 & 60.50 \\
    Change font & 59.46 & 57.89 & 59.46 & 18.92 & 55.26 & 51.35 \\
    \rowcolor{tblZebra} Change layout & 45.20 & 50.03 & 43.40 & 16.03 & 55.11 & 45.07 \\
    Change theme & 40.61 & 39.56 & 41.52 & 6.56 & 38.46 & 38.80 \\
    \rowcolor{tblZebra} Hide/delete & 35.54 & 31.89 & 36.35 & 10.34 & 34.98 & 39.46 \\
    Move & 22.86 & 20.80 & 24.43 & 1.77 & 23.12 & 24.22 \\
    \rowcolor{tblZebra} Region edit & 45.78 & 37.89 & 43.95 & 9.26 & 38.79 & 45.98 \\
    Replace image & 40.00 & 57.88 & 38.76 & 15.28 & 69.04 & 44.27 \\
    \rowcolor{tblZebra} Revert & 53.33 & 31.04 & 53.10 & 17.51 & 43.74 & 53.23 \\
    Rewrite text & 44.95 & 42.16 & 44.25 & 25.09 & 48.43 & 49.13 \\
    \rowcolor{tblZebra} Swap & 40.17 & 44.65 & 43.97 & 10.11 & 49.38 & 44.25 \\
    \bottomrule
  \end{tabularx}
  \arrayrulecolor{black}
\end{table}

\textbf{The fixed-executor pipeline improves joint correctness for every operation.} Track~C exceeds Track~A by 3.77--14.77 points across all 12 types. The largest increase is on font changes, although that category contains only 37 steps; among operations with at least 1,000 steps, Revert gains 9.01 points. This breadth indicates that the pipeline advantage is not confined to one editing category.

\textbf{The two leading Omni models exhibit different operation profiles.} Gemini 3.5 Flash~\citep{google2026gemini35flash} is strongest in direct editing for Change Font (59.46), Change Color (57.08), and Revert (53.33). Qwen3.5-Omni-Plus~\citep{team2026qwen35omni} reaches 25.09 only on Rewrite Text in Track~A, but its Track~C score exceeds its Track~A score for all 12 operations, with gains of 22.44--42.30 points. This pattern is consistent with stronger instruction recovery and fixed-executor performance than direct code execution.

\textbf{Recovery-to-execution gaps vary across operations.} The Track~A and Track~C operation rankings are strongly aligned ($\rho=0.958$), and Move is the lowest-scoring operation on every track (11.89, 16.03, and 16.88 joint accuracy on A, B, and C). The largest gaps occur for Replace Image (47.26 on B versus 29.42 on C), Layout (41.43 versus 33.05), and Swap (32.03 versus 26.80). For these operations, models receive target-and-action credit more often than the downstream pages receive joint correctness credit.

\begin{figure}[t]
  \centering
  \includegraphics[width=\textwidth]{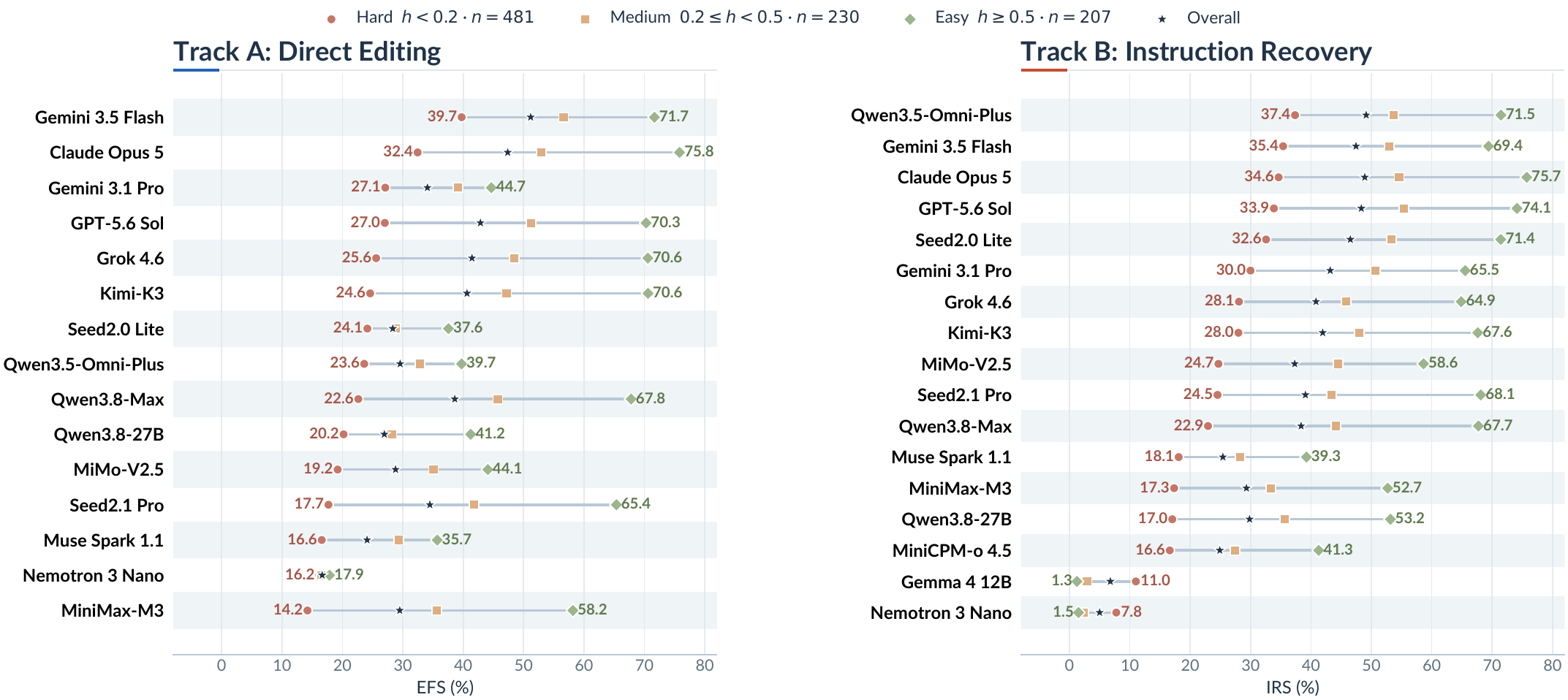}
  \caption{All-model performance across probe-defined visual-dependence strata. Lines connect each model's Hard, Medium, and Easy scores; stars mark overall averages.}
  \label{fig:difficulty-diagnostics}
\end{figure}

\subsection{Controlled Modality and Temporal Alignment Ablations}
\label{app:modality-ablation}

\textbf{Input modality.} Table~\ref{tab:input-modality-ablation} summarizes the aggregate results in the main paper; Tables~\ref{tab:input-modality-ablation-full-ab} and~\ref{tab:input-modality-ablation-full-c} report every component score. Every row uses the same 918 instances and Gemini 3.1 Pro~\citep{google2026gemini31pro} judge. The transcript-only condition supplies the manual step-aligned transcript but no video; the silent-video condition supplies the video without its audio channel together with the transcript; and the audio-video condition supplies the original audio-bearing video without an injected transcript. Track~C changes only the input used to recover instructions and retains DeepSeek-v4-pro~\citep{deepseek2026deepseek} as the fixed executor.

\begin{table}[H]
  \centering
  \scriptsize
  \setlength{\tabcolsep}{2.0pt}
  \caption{Full controlled input-modality results for Tracks A and B on the 918 instances. Input abbreviations are T, V+T, and A+V. Scores are percentages; bold marks each model's best condition.}
  \label{tab:input-modality-ablation-full-ab}
  \arrayrulecolor{tblNavy}
  \begin{tabularx}{\textwidth}{@{}>{\raggedright\arraybackslash}p{2.45cm}c*{5}{Y}!{\hspace{2pt}{\color{tblNavy}\vrule width 0.3pt}\hspace{2pt}}*{3}{Y}@{}}
    \toprule
    \tblHead{Model} & \tblHead{Input} & \tblTrackHead{A}{A Grd.} & \tblTrackHead{A}{A Edit} & \tblTrackHead{A}{A No dmg.} & \tblTrackHead{A}{A Rev.} & \tblTrackHead{A}{A EFS} & \tblTrackHead{B}{B Target F1} & \tblTrackHead{B}{B Action F1} & \tblTrackHead{B}{B IRS} \\
    \midrule
    \modelicon{1.0em}{gemini-color.png}~Gemini 3.5 Flash & T & 30.88 & 20.50 & 49.35 & 47.77 & 32.76 & 4.90 & 31.46 & 18.18 \\
    \rowcolor{tblZebra} & V+T & 37.04 & 23.48 & 41.55 & 53.10 & 34.37 & 30.90 & 45.23 & 38.06 \\
     & A+V & \textbf{59.20} & \textbf{40.52} & \textbf{53.71} & \textbf{71.26} & \textbf{51.17} & \textbf{47.04} & \textbf{47.88} & \textbf{47.46} \\
    \midrule
    \modelicon{1.0em}{qwen-color.png}~Qwen3.5-Omni-Plus & T & 14.97 & 6.51 & 64.43 & 29.76 & 23.59 & 8.46 & \textbf{48.85} & 28.65 \\
    \rowcolor{tblZebra} & V+T & 13.70 & 6.64 & 66.98 & 21.71 & 22.43 & 23.77 & 41.39 & 32.58 \\
     & A+V & \textbf{20.79} & \textbf{11.86} & \textbf{75.46} & \textbf{32.28} & \textbf{29.56} & \textbf{49.99} & 48.29 & \textbf{49.14} \\
    \bottomrule
  \end{tabularx}
  \arrayrulecolor{black}
\end{table}

\begin{table}[H]
  \centering
  \scriptsize
  \setlength{\tabcolsep}{2.0pt}
  \caption{Full controlled input-modality results for Track C on the 918 instances. Input abbreviations are T, V+T, and A+V. Scores are percentages; bold marks each model's best condition.}
  \label{tab:input-modality-ablation-full-c}
  \arrayrulecolor{tblNavy}
  \begin{tabularx}{\textwidth}{@{}>{\raggedright\arraybackslash}p{2.45cm}c*{5}{Y}@{}}
    \toprule
    \tblHead{Model} & \tblHead{Input} & \tblTrackHead{C}{Grd.} & \tblTrackHead{C}{Edit} & \tblTrackHead{C}{No dmg.} & \tblTrackHead{C}{Rev.} & \tblTrackHead{C}{EFS} \\
    \midrule
    \modelicon{1.0em}{gemini-color.png}~Gemini 3.5 Flash & T & 15.06 & 7.18 & 39.04 & 25.45 & 18.00 \\
    \rowcolor{tblZebra} & V+T & 41.59 & 24.71 & 41.02 & 51.23 & 34.96 \\
     & A+V & \textbf{59.16} & \textbf{39.58} & \textbf{51.42} & \textbf{68.92} & \textbf{49.77} \\
    \midrule
    \modelicon{1.0em}{qwen-color.png}~Qwen3.5-Omni-Plus & T & 30.39 & 18.81 & 47.97 & 45.17 & 31.07 \\
    \rowcolor{tblZebra} & V+T & 33.56 & 19.48 & 40.31 & 50.53 & 31.26 \\
     & A+V & \textbf{61.39} & \textbf{41.72} & \textbf{53.06} & \textbf{67.36} & \textbf{51.08} \\
    \bottomrule
  \end{tabularx}
  \arrayrulecolor{black}
\end{table}

\textbf{Native audiovisual input is strongest for both tested Omni models.} It has the highest score in all three tracks among the tested input conditions, supporting the value of preserving synchronized speech and visual context.

\phantomsection
\label{app:temporal-alignment}

\textbf{Temporal alignment.} To isolate temporal alignment from modality content, we sample 300 benchmark instances without using model predictions, scores, or judge decisions. The sample spans all 129 source pages, both languages, all three difficulty bands, all 12 operations, and 4,554 reference steps. For each instance, we circularly shift the audio by five seconds while leaving the video stream unchanged; 150 use a $-5$-second shift and 150 use a $+5$-second shift, preserving all speech content. We evaluate Gemini 3.5 Flash and Qwen3.5-Omni-Plus on the aligned and shifted recordings with the same Track~B prompt and judge. Confidence intervals use 10,000 paired source-page cluster bootstrap replicates.

\begin{table}[H]
  \centering
  \caption{Track~B temporal-alignment intervention on 300 instances. $\Delta$ is aligned minus shifted IRS; intervals are paired source-page cluster bootstrap 95\% CIs.}
  \label{tab:temporal-alignment-intervention}
  \scriptsize
  \setlength{\tabcolsep}{3pt}
  \renewcommand{\arraystretch}{1.06}
  \arrayrulecolor{tblNavy}
  \begin{tabularx}{\textwidth}{@{}>{\raggedright\arraybackslash}p{3.2cm}*{4}{Y}@{}}
    \toprule
    \tblHead{Model} & \tblTrackHead{B}{Aligned} & \tblTrackHead{B}{Shifted} & \tblHead{$\Delta$} & \tblHead{95\% CI} \\
    \midrule
    \modelicon{1.0em}{gemini-color.png}~Gemini 3.5 Flash & 47.65 & 35.99 & 11.65 & [9.37, 13.92] \\
    \rowcolor{tblZebra}\modelicon{1.0em}{qwen-color.png}~Qwen3.5-Omni-Plus & 48.39 & 34.66 & 13.73 & [11.54, 16.00] \\
    \bottomrule
  \end{tabularx}
  \arrayrulecolor{black}
\end{table}

Temporal misalignment substantially reduces instruction recovery for both models. IRS drops by 11.65 points for Gemini 3.5 Flash and 13.73 points for Qwen3.5-Omni-Plus. Target F1 falls by 18.26 and 20.81 points, respectively, compared with Action F1 decreases of 5.04 and 6.66 points, showing that disrupted timing primarily impairs grounding the spoken request to the referenced page region.

\textbf{Timestamped transcripts.} We further evaluate whether explicit temporal information improves VL instruction recovery on the same fixed 300-instance set. We detect speech intervals from the isolated audio, use Qwen3-ASR-Flash~\citep{shi2026qwen3} to audit the recognized content in each interval, and align the manually annotated step-level transcript with the corresponding intervals. The resulting input preserves the manual transcript verbatim and only prefixes each instruction with its start and end times, such as \texttt{[t=15s--23s]}. Both conditions use the same 1 FPS video frames, transcript text, Track~B prompt, and scoring protocol.

\begin{table}[H]
  \centering
  \small
  \setlength{\tabcolsep}{4pt}
  \renewcommand{\arraystretch}{1.06}
  \caption{Track~B results with plain and timestamped transcripts on the fixed 300-instance subset (300 cases for GPT-5.6 Sol and 298 completed cases for Grok 4.6). Plain+F denotes the plain manual transcript with 1 FPS frames; TS+F denotes the timestamped manual transcript with the same frames. Gain is TS+F minus Plain+F.}
  \label{tab:timestamped-transcript-ablation}
  \arrayrulecolor{tblNavy}
  \begin{tabularx}{\textwidth}{@{}>{\raggedright\arraybackslash}p{3.0cm}>{\raggedright\arraybackslash}p{2.4cm}*{3}{Y}@{}}
    \toprule
    \tblHead{Model} & \tblHead{Input} & \tblTrackHead{B}{Target F1} & \tblTrackHead{B}{Action F1} & \tblTrackHead{B}{IRS} \\
    \midrule
    \modelicon{1.0em}{openai.png}~GPT-5.6 Sol & Plain+F & 48.29 & 49.49 & 48.89 \\
    \rowcolor{tblZebra} & TS+F & \textbf{59.31} & \textbf{55.99} & \textbf{57.65} \\
    & Gain & \textbf{+11.02} & \textbf{+6.50} & \textbf{+8.76} \\
    \midrule
    \rowcolor{tblZebra}\modelicon{1.0em}{grok.png}~Grok 4.6 & Plain+F & 33.23 & 47.71 & 40.47 \\
    & TS+F & \textbf{56.68} & \textbf{54.23} & \textbf{55.45} \\
    \rowcolor{tblZebra} & Gain & \textbf{+23.45} & \textbf{+6.52} & \textbf{+14.98} \\
    \bottomrule
  \end{tabularx}
  \arrayrulecolor{black}
\end{table}

\textbf{Explicit temporal alignment primarily improves target grounding for VL models.} For GPT-5.6 Sol~\citep{openai2026gpt56}, timestamps raise IRS by 8.76 points, with a larger gain in Target F1 (+11.02) than Action F1 (+6.50); the paired source-page cluster bootstrap gives a 95\% confidence interval of [6.55, 11.04] for the IRS gain. Together with the degradation under shifted audio, this result shows that temporal correspondence between speech and page state is a central task signal, beyond the speech content alone.

\textbf{The main leaderboard measures end-to-end performance under native model interfaces.} We do not provide timestamp augmentation to VL models in the main results because constructing it requires separate audio processing and an external ASR system, whose capability is not native to the evaluated VL model. Adding this information instead evaluates an augmented alignment-plus-VL pipeline. The main results therefore use the plain manual transcript with visual input, while this controlled experiment shows that an external temporal-alignment module can substantially improve VL instruction recovery.

\subsection{Same-Model Execution Control}
\label{app:self-execution-control}

Track~C uses a fixed DeepSeek-v4-pro code executor, so its recovered-to-oracle gap may depend on the executor choice. To test whether the gap is solely caused by the external executor, we reuse the fixed 300-instance sample from Appendix~\ref{app:temporal-alignment} and evaluate Gemini 3.5 Flash and Qwen3.5-Omni-Plus with same-model execution.

For each instruction-recovery model, we compare two instruction sources and two executors. The first source is the editing instructions recovered by the model from the original audiovisual recording; the second is the benchmark's human-authored oracle instructions. Each instruction set is executed both by the fixed DeepSeek-v4-pro executor and by the instruction-recovery model itself, using the same source HTML, executor prompt, output budget, and EFS protocol within each comparison. The reported EFS therefore evaluates the final HTML generated after executing the corresponding instructions, rather than the instruction text itself. Missing or unusable outputs follow the main zero-fill policy.

\begin{table*}[t]
  \centering
  \scriptsize
  \setlength{\tabcolsep}{3.0pt}
  \renewcommand{\arraystretch}{1.06}
  \caption{Execution control on the same 300 instances. Recovered and Oracle report the EFS of the final HTML generated from model-recovered and benchmark-reference instructions, respectively. Gap is Oracle minus Recovered; Ratio is Recovered divided by Oracle.}
  \label{tab:self-execution-control}
  \arrayrulecolor{tblNavy}
  \begin{tabularx}{\textwidth}{@{}>{\raggedright\arraybackslash}p{3.2cm}>{\raggedright\arraybackslash}p{2.7cm}*{4}{Y}@{}}
    \toprule
    \tblHead{Instruction model} & \tblHead{Executor} & \tblHead{Recovered} & \tblHead{Oracle} & \tblHead{Gap} & \tblHead{Ratio} \\
    \midrule
    \modelicon{1.0em}{gemini-color.png}~Gemini 3.5 Flash & DeepSeek-v4-pro & 49.32 & 89.35 & 40.03 & 55.20\% \\
    \rowcolor{tblZebra} & Gemini 3.5 Flash & 53.48 & 90.45 & 36.96 & 59.13\% \\
    \midrule
    \modelicon{1.0em}{qwen-color.png}~Qwen3.5-Omni-Plus & DeepSeek-v4-pro & 50.32 & 89.35 & 39.03 & 56.32\% \\
    \rowcolor{tblZebra} & Qwen3.5-Omni-Plus & 45.20 & 67.43 & 22.24 & 67.02\% \\
    \bottomrule
  \end{tabularx}
  \arrayrulecolor{black}
\end{table*}

\textbf{Executor choice alone does not explain the recovered-to-oracle gap.} For Gemini 3.5 Flash, self-execution raises recovered-instruction EFS from 49.32 to 53.48 and oracle-instruction EFS from 89.35 to 90.45, leaving a 36.96-point gap. For Qwen3.5-Omni-Plus, the scores for recovered and oracle instructions are 50.32 and 89.35 with DeepSeek-v4-pro, and 45.20 and 67.43 with self-execution, leaving gaps of 39.03 and 22.24 points. Substantial gaps persist under both executors, but this control does not by itself attribute the remaining gap to understanding or execution.

\subsection{Alternative-Judge Ablation}
\label{app:judge-ablation}

The alternative-judge analysis uses a fixed 300-instance evaluator-validation set, with 100 instances from each visual-dependence band under seed 20260824~\citep{zheng2023judging,panickssery2024llm}. Each judge evaluates the same 300-instance grid and 4,482 reference steps under the zero-fill policy. The selected judge is used throughout both the rubric and No-damage stages; the experiment does not replace only one component of the scoring pipeline. Table~\ref{tab:judge-robustness-full} reports all component scores underlying the summary in Table~\ref{tab:judge-robustness}.

% Source SHA256: f6e0e58e0980575abe37d6f614d8a8e3dfffb5f8aa22a3516327c174170deda6
\begin{table}[H]
  \centering
  \scriptsize
  \setlength{\tabcolsep}{3.0pt}
  \renewcommand{\arraystretch}{1.06}
  \caption{Full Track~A judge-ablation results on the same 300 instances. Scores are percentages; parentheses give each model's rank under that judge for the corresponding column.}
  \label{tab:judge-robustness-full}
  \arrayrulecolor{tblNavy}
  \begin{tabularx}{\textwidth}{@{}>{\raggedright\arraybackslash}p{2.35cm}>{\raggedright\arraybackslash}p{2.0cm}*{5}{Y}@{}}
    \toprule
    \tblHead{Model} & \tblHead{Judge} & \tblTrackHead{A}{Grd.} & \tblTrackHead{A}{Edit} & \tblTrackHead{A}{No dmg.} & \tblTrackHead{A}{Rev.} & \tblTrackHead{A}{EFS} \\
    \midrule
    \modelicon{1.0em}{gemini-color.png}~Gemini 3.5 Flash & Gemini 3.1 Pro & 62.78 (2) & \textbf{36.58 (1)} & 60.95 (4) & \textbf{73.39 (1)} & \textbf{51.44 (1)} \\
    \rowcolor{tblZebra}  & GPT-5.5 & 56.94 (2) & 32.67 (2) & 67.00 (4) & 76.44 (2) & \textbf{50.72 (1)} \\
     & DeepSeek-v4-pro & 52.01 (2) & 33.88 (2) & 65.43 (4) & \textbf{70.11 (1)} & \textbf{49.29 (1)} \\
    \midrule
    \modelicon{1.0em}{openai.png}~GPT-5.6 Sol & Gemini 3.1 Pro & \textbf{64.03 (1)} & 36.21 (2) & 32.95 (6) & 70.89 (2) & 45.28 (2) \\
    \rowcolor{tblZebra}  & GPT-5.5 & \textbf{57.61 (1)} & \textbf{32.77 (1)} & 40.88 (6) & 69.39 (6) & 44.20 (3) \\
     & DeepSeek-v4-pro & \textbf{54.83 (1)} & \textbf{34.06 (1)} & 38.33 (6) & 67.61 (2) & 43.73 (2) \\
    \midrule
    \modelicon{1.0em}{kimi.png}~Kimi-K3 & Gemini 3.1 Pro & 50.99 (3) & 30.44 (3) & 53.87 (5) & 62.78 (3) & 43.65 (3) \\
    \rowcolor{tblZebra}  & GPT-5.5 & 48.33 (3) & 27.39 (3) & 58.53 (5) & 71.06 (5) & 44.44 (2) \\
     & DeepSeek-v4-pro & 44.30 (3) & 28.48 (3) & 60.12 (5) & 62.78 (3) & 43.38 (3) \\
    \midrule
    \modelicon{1.0em}{gemini-color.png}~Gemini 3.1 Pro & Gemini 3.1 Pro & 28.78 (4) & 15.47 (4) & \textbf{76.92 (1)} & 42.83 (4) & 34.56 (4) \\
    \rowcolor{tblZebra}  & GPT-5.5 & 30.62 (4) & 14.42 (4) & \textbf{79.11 (1)} & \textbf{77.94 (1)} & 41.68 (4) \\
     & DeepSeek-v4-pro & 23.53 (4) & 15.02 (4) & \textbf{79.07 (1)} & 55.44 (4) & 36.81 (4) \\
    \midrule
    \modelicon{1.0em}{qwen-color.png}~Qwen3.5-Omni-Plus & Gemini 3.1 Pro & 27.62 (5) & 14.04 (5) & 74.77 (2) & 38.06 (5) & 32.35 (5) \\
    \rowcolor{tblZebra}  & GPT-5.5 & 26.33 (5) & 13.05 (5) & 77.47 (2) & 73.28 (4) & 39.31 (5) \\
     & DeepSeek-v4-pro & 21.59 (5) & 13.51 (5) & 78.84 (2) & 46.56 (5) & 34.01 (5) \\
    \midrule
    \modelicon{1.0em}{nvidia.png}~Nemotron 3 Nano & Gemini 3.1 Pro & 1.06 (6) & 0.00 (6) & 72.10 (3) & 9.44 (6) & 16.42 (6) \\
    \rowcolor{tblZebra}  & GPT-5.5 & 6.61 (6) & 0.42 (6) & 74.08 (3) & 75.22 (3) & 30.73 (6) \\
     & DeepSeek-v4-pro & 1.52 (6) & 0.45 (6) & 73.78 (3) & 29.83 (6) & 21.10 (6) \\
    \bottomrule
  \end{tabularx}
  \arrayrulecolor{black}
\end{table}

\subsection{Human Calibration of the LLM Judge}
\label{app:human-judge-calibration}

We constructed a paired human-calibration pool from 50 benchmark instances, with one Track~A and one Track~B output per instance. Sampling used fixed seeds and did not use model scores, rankings, or judge decisions. The pool balances four model families: in the order Claude Opus 5~\citep{anthropic2026claudeopus5}, Gemini 3.5 Flash~\citep{google2026gemini35flash}, Qwen3.5-Omni-Plus~\citep{team2026qwen35omni}, and Qwen3.8-27B~\citep{qwen2026qwen38}, the Track~A allocation is 12/13/12/13 and the Track~B allocation is 13/13/12/12. Its 50 instances come from 50 distinct source pages and comprise 24 English and 26 Chinese examples, with 12 Easy, 12 Medium, and 26 Hard. All 12 operations are represented: eight common types occur in all 50 source histories, while Hide/Delete, Change Color, Rewrite Text, and Change Font occur in 49, 37, 15, and 3, respectively. The only availability requirement was that the assigned model output had already been generated and could be inspected.

The human evaluator inspected the source page, request, and model output, judging target grounding and edit or revision fulfillment for Track~A, and target/action recovery plus step alignment for Track~B. Automatic decisions remained hidden until the corresponding human decisions were submitted, and all analyses use the first answers frozen before judge reveal. The evaluator completed all the outputs, covering 1,451 steps and 2,902 valid binary decisions: 653 steps for Track~A and 798 for Track~B. This calibration targets the rubric components decided by the LLM judge; the deterministic No-damage component is outside its scope.

\begin{table}[t]
  \centering
  \caption{Blind human calibration of the automatic judge. Agreement and Cohen's $\kappa$~\citep{cohen1960coefficient} are computed over binary item decisions; Pearson $r$ and bias are computed over output-level component scores. Bias is automatic minus human score in percentage points.}
  \label{tab:human-judge-calibration}
  \small
  \setlength{\tabcolsep}{4pt}
  \renewcommand{\arraystretch}{1.06}
  \arrayrulecolor{tblNavy}
  \begin{tabularx}{\textwidth}{@{}>{\raggedright\arraybackslash}p{0.8cm}>{\raggedright\arraybackslash}p{2.5cm}*{5}{Y}@{}}
    \toprule
    \tblHead{Track} & \tblHead{Component} & \tblHead{Items} & \tblHead{Agreement} & \tblHead{$\kappa$} & \tblHead{Pearson $r$} & \tblHead{Bias} \\
    \midrule
    \tblTrackLabel{A} & Target grounding  & 650 & 75.69\% & 0.502 & 0.867 & $-1.00$ \\
    \rowcolor{tblZebra} \tblTrackLabel{A} & Edit fulfillment  & 528 & 76.52\% & 0.459 & 0.840 & $-10.65$ \\
    \tblTrackLabel{A} & Revision          & 122 & 68.85\% & 0.337 & 0.361 & $\phantom{-}0.00$ \\
    \midrule
    \tblTrackLabel{B} & Target-F1         & 798 & 81.58\% & 0.630 & 0.888 & $-10.16$ \\
    \rowcolor{tblZebra} \tblTrackLabel{B} & Action-F1         & 798 & 80.14\% & 0.606 & 0.851 & $-8.60$ \\
    \tblTrackLabel{B} & IRS               & --- & --- & --- & \textbf{0.906} & $-9.38$ \\
    \bottomrule
  \end{tabularx}
  \arrayrulecolor{black}
\end{table}

Human and automatic judgments are broadly aligned. On Track~A, output-level correlations reach $r=0.867$ for grounding and $r=0.840$ for edit fulfillment. Agreement is higher on Track~B: Target and Action exceed 80\% item agreement, step alignment agrees in 91.39\% of cases, and IRS reaches $r=0.906$. The judge is more conservative on edit fulfillment and Track~B, but preserves relative variation across outputs. Together with the alternative-judge results, this human check supports using the automatic evaluator as a consistent and reproducible basis for model comparison.

\subsection{EFS Hyperparameter Sensitivity}
\label{app:efs-sensitivity}

We evaluate whether the EFS rankings depend on its category weights or the No-damage normalization coefficient $\alpha$. The sweep varies $\alpha\in\{0.05,0.10,0.15,0.20,0.25,0.30\}$ and uses 26 weight vectors that sum to one, including a balanced vector and configurations in which Edit Fulfillment has the largest or tied-largest weight. Their Cartesian product gives 156 configurations; the baseline is $(w_G,w_E,w_N,w_R)=(0.10,0.50,0.20,0.20)$ with $\alpha=0.15$. All configurations reuse the same predictions and judgments and retain the primary zero-fill policy.

\begin{table}[H]
  \centering
  \scriptsize
  \setlength{\tabcolsep}{3.0pt}
  \renewcommand{\arraystretch}{1.06}
  \caption{Ranking sensitivity to EFS weights and the No-damage coefficient $\alpha$. Spearman $\rho$ compares each configuration with the baseline ranking.}
  \label{tab:efs-sensitivity}
  \arrayrulecolor{tblNavy}
  \begin{tabularx}{\textwidth}{@{}>{\raggedright\arraybackslash}p{0.8cm}>{\raggedright\arraybackslash}p{2.4cm}*{5}{Y}@{}}
    \toprule
    \tblHead{Track} & \tblHead{Variation} & \tblHead{Configs} & \tblHead{Leader stable} & \tblHead{Median $\rho$} & \tblHead{Min. $\rho$} & \tblHead{Max. rank shift} \\
    \midrule
    \tblTrackLabel{A} & Weights only & 26 & 26/26 & 0.982 & 0.743 & 6 \\
    \rowcolor{tblZebra} \tblTrackLabel{A} & $\alpha$ only & 6 & 6/6 & 0.993 & 0.986 & 2 \\
    \tblTrackLabel{A} & Joint & 156 & 156/156 & 0.977 & 0.639 & 8 \\
    \midrule
    \tblTrackLabel{C} & Weights only & 26 & 26/26 & 0.998 & 0.988 & 2 \\
    \rowcolor{tblZebra} \tblTrackLabel{C} & $\alpha$ only & 6 & 6/6 & 0.999 & 0.995 & 1 \\
    \tblTrackLabel{C} & Joint & 156 & 156/156 & 0.998 & 0.971 & 3 \\
    \bottomrule
  \end{tabularx}
  \arrayrulecolor{black}
\end{table}

\textbf{The top-ranked models are stable across the tested configurations.} Gemini 3.5 Flash remains first on Track~A, while Qwen3.5-Omni-Plus remains first on Track~C. Varying only $\alpha$ has little effect: the minimum Spearman correlation is $0.986$ on Track~A and $0.995$ on Track~C. Rank changes beyond the leader occur more often, so we treat the top model as stable within this sweep without claiming that the full ranking is invariant.

\textbf{Weight sensitivity is concentrated in Track~A's middle and lower ranks.} Track~C remains highly stable under weight-only changes, with a minimum Spearman correlation of $0.988$. On Track~A, the No-damage-heavy vector $(0.10,0.40,0.40,0.10)$ lowers $\rho$ to $0.743$ and moves Qwen3.8-27B from rank 13 to rank 7. Combining this vector with $\alpha=0.05$ gives the joint minimum $\rho=0.639$; nevertheless, 150 of 156 joint configurations retain $\rho\geq0.9$, and none changes the leader. Thus, the top-ranked models are stable across the tested EFS parameterizations, while Track~A's full ordering reflects the metric's tradeoff between preservation and requested edits.

\subsection{Track B Metric Sensitivity}
\label{app:track-b-sensitivity}

The primary IRS averages target and action F1, so unsupported predicted steps reduce precision without changing recall. For component $c\in\{T,A\}$, let
{\small
\begin{gather}
  P_i^c=\frac{m_i^c}{n_i^{\mathrm{pred}}},
  \qquad
  R_i^c=\frac{m_i^c}{n_i^{\mathrm{ref}}}, \\
  F_{\beta,i}^c=\frac{(1+\beta^2)P_i^cR_i^c}{\beta^2P_i^c+R_i^c}, \\
  \mathrm{IRS}_{\beta}=\frac{1}{2N}\sum_{i=1}^{N}
  \left(F_{\beta,i}^{T}+F_{\beta,i}^{A}\right).
\end{gather}}%
Here $N=918$. We set $P_i^c=0$ when no steps are predicted and $F_{\beta,i}^c=0$ whenever $P_i^cR_i^c=0$. The primary metric uses $\beta=1$; $F_{0.5}$ emphasizes precision and therefore penalizes unsupported predictions more strongly, whereas $F_2$ emphasizes recall. We compare these variants with recall-only scoring and a stricter Joint-F1 that credits a step only when both its target and action are correct.

\begin{table}[H]
  \centering
  \scriptsize
  \setlength{\tabcolsep}{3pt}
  \renewcommand{\arraystretch}{1.06}
  \caption{Track~B ranking sensitivity. Correlations and rank shifts are measured against the primary Component-F1 IRS ranking.}
  \label{tab:track-b-metric-sensitivity}
  \arrayrulecolor{tblNavy}
  \begin{tabularx}{\textwidth}{@{}>{\raggedright\arraybackslash}p{2.8cm}>{\raggedright\arraybackslash}p{3.4cm}*{3}{Y}@{}}
    \toprule
    \tblHead{Metric} & \tblHead{Leader} & \tblHead{$\rho$} & \tblHead{Max shift} & \tblHead{Same top 3} \\
    \midrule
    Component F1 (primary) & \modelicon{0.9em}{qwen-color.png}~Qwen3.5-Omni-Plus & 1.000 & 0 & Yes \\
    \rowcolor{tblZebra} Recall only & \modelicon{0.9em}{qwen-color.png}~Qwen3.5-Omni-Plus & 0.998 & 1 & Yes \\
    Component $F_{0.5}$ & \modelicon{0.9em}{gemini-color.png}~Gemini 3.5 Flash & 0.983 & 3 & No \\
    \rowcolor{tblZebra} Component $F_2$ & \modelicon{0.9em}{qwen-color.png}~Qwen3.5-Omni-Plus & 0.998 & 1 & Yes \\
    Joint F1 & \modelicon{0.9em}{claude-color.png}~Claude Opus 5 & 0.983 & 2 & Yes \\
    \bottomrule
  \end{tabularx}
  \arrayrulecolor{black}
\end{table}

\textbf{The broad ranking is stable, but the exact leader depends on how precision is weighted.} Every alternative has Spearman $\rho\geq0.983$ relative to the primary ranking. Qwen3.5-Omni-Plus remains first under recall-only and $F_2$, whereas $F_{0.5}$ favors the more concise outputs of Gemini 3.5 Flash and Joint-F1 places Claude Opus 5 first. The primary Component-F1 definition balances recovered coverage with unsupported-output penalties without requiring target and action to pass jointly.

\subsection{Source-Page Cluster Bootstrap}
\label{app:source-page-bootstrap}

The 918 instances derive from 129 source webpages, and instances from the same page reuse its layout, components, and visual style. They may therefore be correlated rather than independent observations. Treating all instances as independent could overstate the effective sample size and make model differences appear more stable than they are across distinct webpages. We use a source-page cluster bootstrap~\citep{field2007bootstrapping} to test whether track leaders and the Track~A-to-C changes persist across alternative compositions of source webpages, rather than being driven by a small number of pages.

Each replicate samples 129 page IDs with replacement. A page drawn $k$ times contributes all of its instances $k$ times, while unselected pages contribute none. The same draw is used for every model and track, preserving paired comparisons. Across 10,000 replicates, we recompute instance-macro EFS or IRS, ranks, pairwise gaps, and C--A differences. In Table~\ref{tab:source-page-bootstrap}, $p_+$ is the fraction of positive differences; ``Stable $+$'' or ``Stable $-$'' requires at least 97.5\% in the corresponding direction.

\begin{table}[H]
  \centering
  \small
  \setlength{\tabcolsep}{4pt}
  \renewcommand{\arraystretch}{1.06}
  \caption{Selected source-page cluster bootstrap comparisons over 10,000 resamples. $\Delta$ is the first score minus the second, or Track~C minus Track~A; $p_+$ is the percentage of positive resamples.}
  \label{tab:source-page-bootstrap}
  \arrayrulecolor{tblNavy}
  \begin{tabularx}{\columnwidth}{@{}>{\raggedright\arraybackslash}X>{\centering\arraybackslash}p{1.1cm}>{\centering\arraybackslash}p{1.5cm}>{\centering\arraybackslash}p{1.6cm}@{}}
    \toprule
    \tblHead{Comparison} & \tblHead{$\Delta$} & \tblHead{$p_+$} & \tblHead{Direction} \\
    \midrule
    Track A: Gemini 3.5 Flash - Claude Opus 5 & +3.80 & 99.99\% & Stable + \\
    \rowcolor{tblZebra} Track B: Qwen3.5-Omni-Plus - Claude Opus 5 & +0.23 & 60.30\% & Uncertain \\
    Track C: Qwen3.5-Omni-Plus - Gemini 3.5 Flash & +1.31 & 98.63\% & Stable + \\
    \midrule
    A$\rightarrow$C: Seed2.0 Lite & +19.09 & 100.00\% & Stable + \\
    \rowcolor{tblZebra} A$\rightarrow$C: Gemini 3.5 Flash & -1.40 & 1.09\% & Stable - \\
    A$\rightarrow$C: Qwen3.5-Omni-Plus & +21.52 & 100.00\% & Stable + \\
    \rowcolor{tblZebra} A$\rightarrow$C: Grok 4.6 & -2.85 & 0.00\% & Stable - \\
    \bottomrule
  \end{tabularx}
  \arrayrulecolor{black}
\end{table}

\textbf{The Direct Editing leader is stable, whereas the leading Track~B pair remains close.} Gemini 3.5 Flash ranks first on Track~A in 99.99\% of resamples. Qwen3.5-Omni-Plus ranks first on Track~B in 55.67\% of resamples, and its 0.23-point lead over Claude Opus 5 has uncertain direction. It retains the highest Track~C EFS in 98.63\% of resamples; its 1.31-point lead over Gemini 3.5 Flash is positive in 98.63\% of the paired resamples.

\textbf{The pipeline benefit is model dependent.} Of the 15 models with both Track~A and Track~C results, six have positive C--A differences in at least 97.5\% of resamples and two have a negative difference at the same threshold; the remaining seven have uncertain direction. The large gains for Qwen3.5-Omni-Plus and Seed2.0 Lite remain positive in every resample, whereas Grok 4.6 consistently favors direct editing. As a separate sensitivity check, averaging pages equally preserves the Track~A, B, and C leaders, with Spearman correlations of $0.982$, $1.000$, and $0.995$ against the instance-macro rankings.

\subsection{Coverage and Zero-Fill Robustness}
\label{app:coverage-robustness}

\textbf{Zero-filled failures do not determine the leading models.} The primary aggregation retains all 918 instances and assigns zero to refused, empty, unparsable, or otherwise unusable outputs. When we recompute each model's score over valid outputs only, all three tracks retain their leaders and the Track~B top-three set remains unchanged. No available model--track contains more than 41 zero-filled instances. Table~\ref{tab:model-coverage-full} reports the complete per-model coverage.

\begin{table}[H]
  \centering
  \small
  \setlength{\tabcolsep}{4pt}
  \renewcommand{\arraystretch}{1.06}
  \caption{Robustness to zero-filled unusable outputs on the 918 instances. Valid-only scores omit non-valid outputs.}
  \label{tab:coverage-robustness}
  \arrayrulecolor{tblNavy}
  \begin{tabularx}{\textwidth}{@{}>{\raggedright\arraybackslash}X*{5}{Y}@{}}
    \toprule
    \tblHead{Track} & \tblHead{Models} & \tblHead{Affected} & \tblHead{Zero-filled} & \textcolor{tblNavy}{$\boldsymbol{\rho}$} & \textcolor{tblNavy}{$\boldsymbol{\Delta_{\max}}$} \\
    \midrule
    \tblTrackLabel{A} & 15 & 6 & 133 & 0.989 & 2.21 \\
    \rowcolor{tblZebra} \tblTrackLabel{B} & 17 & 8 & 87 & 0.998 & 1.30 \\
    \tblTrackLabel{C} & 17 & 8 & 87 & 0.998 & 1.27 \\
    \bottomrule
  \end{tabularx}
  \arrayrulecolor{black}
\end{table}

\begin{table}[H]
  \centering
  \scriptsize
  \caption{Per-model output coverage. Entries are valid / zero-filled outputs out of 918; ``--'' denotes unavailable Track~A runs.}
  \label{tab:model-coverage-full}
  \arrayrulecolor{tblNavy}
  \begin{tabularx}{\columnwidth}{@{}>{\raggedright\arraybackslash}Xccc@{}}
    \toprule
    \tblHead{Model} & \tblTrackHead{A}{Track A} & \tblTrackHead{B}{Track B} & \tblTrackHead{C}{Track C} \\
    \midrule
    \modelicon{0.9em}{minicpm.png}~MiniCPM-o 4.5 & -- & 918 / 0 & 918 / 0 \\
    \rowcolor{tblZebra}\modelicon{0.9em}{gemini-color.png}~Gemma 4 12B & -- & 913 / 5 & 913 / 5 \\
    \modelicon{0.9em}{qwen-color.png}~Qwen3.8-27B & 918 / 0 & 917 / 1 & 917 / 1 \\
    \rowcolor{tblZebra}\modelicon{0.9em}{nvidia.png}~Nemotron 3 Nano & 918 / 0 & 918 / 0 & 918 / 0 \\
    \modelicon{0.9em}{xiaomimimo.png}~MiMo-V2.5 & 893 / 25 & 915 / 3 & 915 / 3 \\
    \rowcolor{tblZebra}\modelicon{0.9em}{minimax-color.png}~MiniMax-M3 & 881 / 37 & 879 / 39 & 879 / 39 \\
    \modelicon{0.9em}{qwen-color.png}~Qwen3.8-Max & 918 / 0 & 918 / 0 & 918 / 0 \\
    \rowcolor{tblZebra}\modelicon{0.9em}{kimi.png}~Kimi-K3 & 915 / 3 & 902 / 16 & 902 / 16 \\
    \modelicon{0.9em}{meta.png}~Muse Spark 1.1 & 918 / 0 & 918 / 0 & 918 / 0 \\
    \rowcolor{tblZebra}\modelicon{0.9em}{doubao-color.png}~Seed2.0 Lite & 910 / 8 & 911 / 7 & 911 / 7 \\
    \modelicon{0.9em}{gemini-color.png}~Gemini 3.5 Flash & 918 / 0 & 918 / 0 & 918 / 0 \\
    \rowcolor{tblZebra}\modelicon{0.9em}{qwen-color.png}~Qwen3.5-Omni-Plus & 918 / 0 & 918 / 0 & 918 / 0 \\
    \modelicon{0.9em}{doubao-color.png}~Seed2.1 Pro & 918 / 0 & 918 / 0 & 918 / 0 \\
    \rowcolor{tblZebra}\modelicon{0.9em}{grok.png}~Grok 4.6 & 899 / 19 & 906 / 12 & 907 / 11 \\
    \modelicon{0.9em}{gemini-color.png}~Gemini 3.1 Pro & 918 / 0 & 918 / 0 & 918 / 0 \\
    \rowcolor{tblZebra}\modelicon{0.9em}{openai.png}~GPT-5.6 Sol & 918 / 0 & 918 / 0 & 918 / 0 \\
    \modelicon{0.9em}{claude-color.png}~Claude Opus 5 & 877 / 41 & 914 / 4 & 913 / 5 \\
    \bottomrule
  \end{tabularx}
  \arrayrulecolor{black}
\end{table}

\clearpage
\subsection{Prompt Documentation}
\label{app:prompts}

This appendix documents the model-facing prompts used by the evaluation pipeline. Each box contains only text supplied to the model; titles, delivery information, and other documentation remain outside the boxes. Bracketed uppercase text inside a box denotes content substituted at runtime. The boxes reproduce the original prompts used for the reported evaluation.

\definecolor{promptBackground}{HTML}{EDF5FB}
\definecolor{promptAccent}{HTML}{197D80}
\definecolor{promptText}{HTML}{45525D}
\newtcolorbox{promptbody}{%
  enhanced,
  breakable,
  colback=promptBackground,
  coltext=promptText,
  frame hidden,
  borderline west={2.2pt}{0pt}{promptAccent},
  boxrule=0pt,
  sharp corners,
  boxsep=0pt,
  left=3mm,
  right=3mm,
  top=2mm,
  bottom=2mm,
  pad at break*=1.5mm,
  before skip=0pt,
  after skip=10pt
}
\newenvironment{promptbox}[1]{%
  \par\addvspace{10pt}\needspace{100pt}%
  \noindent{\normalfont\bfseries #1}\par\nobreak\vspace{5pt}%
  \begin{promptbody}%
}{\end{promptbody}}
\newcommand{\promptfile}[1]{%
  \VerbatimInput[
    fontsize=\footnotesize,
    baselinestretch=1.06,
    breaklines=true,
    breaksymbolleft={}
  ]{#1}%
}

\subsubsection{Contestant and Executor Prompts}
\label{app:contestant-prompts}

\phantomsection
\label{app:track-a-prompt}
Track~A uses three retained input variants. The \texttt{omni} message accompanies an audio-bearing video; \texttt{vl\_asr} accompanies a silent video and includes its transcript; and \texttt{vl\_asr\_frames} accompanies sampled frames and includes the transcript.

\begin{promptbox}{Track A: \texttt{omni} user-message text}
\promptfile{sections/appendix_prompt_text/01_track_a_direct_editing.txt}
\end{promptbox}

\begin{promptbox}{Track A: \texttt{vl\_asr} user-message text}
\promptfile{sections/appendix_prompt_text/01_track_a_direct_editing_vl_asr.txt}
\end{promptbox}

\begin{promptbox}{Track A: \texttt{vl\_asr\_frames} user-message text}
\promptfile{sections/appendix_prompt_text/01_track_a_direct_editing_vl_asr_frames.txt}
\end{promptbox}

\phantomsection
\label{app:track-b-prompt}
Track~B likewise uses an audio-bearing video for \texttt{omni}, a silent video plus transcript for \texttt{vl\_asr}, and sampled frames plus transcript for \texttt{vl\_asr\_frames}. No source HTML is supplied in the main Track~B protocol.

\begin{promptbox}{Track B: \texttt{omni} user-message text}
\promptfile{sections/appendix_prompt_text/02_track_b_instruction_recovery.txt}
\end{promptbox}

\begin{promptbox}{Track B: \texttt{vl\_asr} user-message text}
\promptfile{sections/appendix_prompt_text/02_track_b_instruction_recovery_vl_asr.txt}
\end{promptbox}

\begin{promptbox}{Track B: \texttt{vl\_asr\_frames} user-message text}
\promptfile{sections/appendix_prompt_text/02_track_b_instruction_recovery_vl_asr_frames.txt}
\end{promptbox}

\phantomsection
\label{app:track-c-prompt}
\newpage
\begin{promptbox}{Track C Fixed-Executor Prompt}
\promptfile{sections/appendix_prompt_text/03_track_c_fixed_executor.txt}
\end{promptbox}

\subsubsection{Scoring Prompts}
\label{app:scoring-prompts}

\phantomsection
\label{app:rubric-judge-prompt}
The following no-video judge prompt begins with the complete ordered history, marking each step as \texttt{ACTIVE} or \texttt{CANCELLED}, followed by the confirmed whole-step revision links. Each history line has the form \texttt{[step $k$] [status] instruction}; cancelled requests provide context only, and the rubric block contains only active items.

\begin{promptbox}{Tracks A and C Rubric-Judge Prompt}
\promptfile{sections/appendix_prompt_text/04_tracks_ac_rubric_judge.txt}
\end{promptbox}

\phantomsection
\label{app:matching-judge-prompt}
\begin{promptbox}{Track B Matching-Judge Prompt}
\promptfile{sections/appendix_prompt_text/05_track_b_matching_judge.txt}
\end{promptbox}

\phantomsection
\label{app:no-damage-prompt}
\begin{promptbox}{No-Damage Judge Prompt}
\promptfile{sections/appendix_prompt_text/06_no_damage_judge.txt}
\end{promptbox}

\phantomsection
\label{app:image-judge-prompt}
The replacement-image judge receives one ordered multimodal user message: the text label \texttt{[ORIGINAL IMAGE]}, the original-image attachment, the text label \texttt{[NEW IMAGE]}, the new-image attachment, and then the instruction text below. If the original image cannot be retrieved, its label and attachment are both omitted.

\begin{promptbox}{Replacement-Image Judge: final text part}
\promptfile{sections/appendix_prompt_text/07_replacement_image_judge.txt}
\end{promptbox}

\subsubsection{Difficulty-Probe Prompts}
\label{app:difficulty-prompts}

The following prompts support the benchmark difficulty analysis.

\phantomsection
\label{app:target-recovery-prompt}
\begin{promptbox}{Text-Only Target-Recovery Prompt}
\promptfile{sections/appendix_prompt_text/08_text_only_target_recovery.txt}
\end{promptbox}

\phantomsection
\label{app:target-equivalence-prompt}
\begin{promptbox}{Target-Equivalence Judge Prompt}
\promptfile{sections/appendix_prompt_text/09_target_equivalence_judge.txt}
\end{promptbox}

\end{document}